\documentclass[format=acmsmall]{acmart}

\usepackage{amsmath,scalerel}
\usepackage{algorithm}
\usepackage{algorithmic}
\usepackage{tabularx, booktabs}
\usepackage{multirow}
\usepackage{slashbox}
\usepackage{amssymb}
\usepackage{pifont}
\usepackage[keeplastbox]{flushend}
\usepackage{float}
\usepackage{siunitx}

\newcolumntype{C}{>{\centering\arraybackslash}p{10ex}}

\DeclareMathOperator*{\argmax}{argmax}
\newcommand{\cmark}{\ding{51}}
\newcommand{\xmark}{\ding{55}}

\acmJournal{TOIS}
\acmVolume{0}
\acmNumber{0}
\acmArticle{7}
\acmYear{2017}
\acmMonth{0}
\copyrightyear{2017}

\setcopyright{acmcopyright}
\acmJournal{TOIS}
\acmYear{2017} \acmVolume{1} \acmNumber{1} \acmArticle{1} \acmMonth{1} 

\acmDOI{0000001.0000001}

\newcommand{\partitle}[1]{\vspace{2mm}\noindent\textbf{#1}}

\graphicspath{{./figures/}}
\begin{document}

\title{Personalized Context-Aware Point of Interest Recommendation}

\author{Mohammad Aliannejadi}
\author{Fabio Crestani}
\affiliation{
\institution{Universit{\`a} della Svizzera italiana (USI)}
\department{Faculty of Informatics}%
\streetaddress{Via Giuseppe Buffi 13}
\city{Lugano}
\country{Switzerland}
\postcode{6900}
}

\begin{abstract}
Personalized recommendation of Points of Interest (POIs) plays a key role in satisfying users on Location-Based Social Networks (LBSNs). In this paper, we propose a probabilistic model to find the mapping between user-annotated tags and locations' taste keywords. Furthermore, we introduce a dataset on locations' contextual appropriateness and demonstrate its usefulness in predicting the contextual relevance of locations. We investigate four approaches to use our proposed mapping for addressing the data sparsity problem: one model to reduce the dimensionality of location taste keywords and three models to predict user tags for a new location. Moreover, we present different scores calculated from multiple LBSNs and show how we incorporate new information from the mapping into a POI recommendation approach. Then, the computed scores are integrated using learning to rank techniques. The experiments on two TREC datasets show the effectiveness of our approach, beating state-of-the-art methods.
\end{abstract}



\thanks{This work is supported by the Swiss National Science Foundation,
  under the project Relevance Criteria Combination for Mobile Information Retrieval (RelMobIR).

  Author's addresses: Mohammad Aliannejadi and Fabio Crestani, Faculty of Informatics, Universit{\`a} della Svizzera italiana (USI), Lugano, Switzerland;
  emails: \{mohammad.alian.nejadi, fabio.crestani\}@usi.ch
  }

\maketitle

\section{Introduction}

Nowadays with the super fast growth in the amount of information available on the world wide web and with the increasing popularity of mobile devices equipped with easy access to the Internet, it is essential to assist users to find relevant and useful information according to their needs and context. Recommender systems aim to filter information in order to satisfy the users' information needs and minimize the effort made by the user to find relevant information. More specifically, recommender systems focus on suggesting items that can potentially be attractive to users~\cite{DBLP:reference/sp/RicciRS15}. 
The availability of \textit{Location-Based Social Networks} (LBSNs) together with a \textit{Global Positioning System} (GPS) and an always-on Internet access on mobile devices encourages many users to check in at various \textit{Points of Interest} (POIs) using their favorite platform. Among the many available LBSNs, we can point out some of the most popular ones. Yelp\footnote{\url{http://www.yelp.com}} is one of the most popular LBSNs in the US and mainly relies on users' reviews.
Users leave very creative and usually long reviews about almost any type of POI on this service. According to the company's factsheet\footnote{\label{yelp-ref}Yelp Inc. 2017. An Introduction to Yelp Metrics as of June 30, 2016. \url{http://web.archive.org/web/20160825213451/https://www.yelp.com/factsheet.} (2017). Accessed: 2017-03-23.}, 73\% of searches on Yelp were performed on a mobile device\footnote{As of June 30, 2016}. Foursquare\footnote{\url{http://www.foursquare.com}} introduces ``tips'' as short informative reviews and ``tastes'' as a way for users to express their expectations and preferences. TripAdvisor\footnote{\url{http://www.tripadvisor.com}}, on the other hand, focuses on reviews and content related to travel and travel companies. The popularity of such services enables them to gather various types of information about users including users' mobility, feedback and context. 
A key factor in satisfying the users' needs is being able to personalize the system to recommend POIs taking into account personal preference and contextual constraints~\cite{DBLP:journals/umuai/ChenCW15}. 

Much work has been carried out to model users taking into consideration their preference history and current vicinity to POIs to help them explore new and interesting locations\footnote{In this paper, we use the terms \textit{location}, \textit{venue} and \textit{POI} interchangeably.}. More specifically, POI recommendation tries to ensure user's satisfaction by suggesting her the most interesting locations, taking into account her preferences and contextual constraints \cite{DBLP:journals/geoinformatica/0003ZWM15,DBLP:journals/umuai/ChenCW15}. In fact, among the many proposed works, there has been several successful approaches \cite{DBLP:conf/uic/ParkHC07,DBLP:conf/kdd/YeSLYJ11,he2016spatial,DBLP:journals/tist/FangXHM16}, addressing different problems of POI recommendation, taking into account various aspects. However, many challenges in this area are not resolved yet.
A major challenge in POI recommendation is data sparsity~\cite{DBLP:journals/tois/YinCZWHS16, DBLP:journals/geoinformatica/0003ZWM15}. Normally, users visit a very limited number of locations, however, LBSNs feature a relatively huge number of locations with a large variety. Consequently, the user-item matrix used in \textit{Collaborative Filtering} (CF) by many papers is sparse~\cite{DBLP:conf/www/ZhengZXY10}. Several studies seek to address the data sparsity problem incorporating additional information into the model~\cite{DBLP:journals/umuai/ChenCW15,DBLP:journals/geoinformatica/0003ZWM15}. 
For instance, \citet{DBLP:journals/tist/ZhangDCLZ13} derived virtual ratings from users' reviews and studied the impact of fusing them into CF. \citet{DBLP:conf/aaai/ZhengCZXY10}, on the other hand, modeled users adopting a tensor representation, and introduced a matrix decomposition and regularized tensor to better address the data sparsity problem. Moreover, since users spend most of their time in their home town~\cite{DBLP:conf/icde/LevandoskiSEM12}, the data sparsity problem is aggravated when a user visits a new city where she has no history of visited locations~\cite{DBLP:conf/cikm/FerenceYL13}. 

Apart from personal preference and interest, the user behavior is influenced and, in many cases, constrained by local and contextual preferences~\cite{DBLP:journals/tist/FangXHM16}. For instance, a user may be a big fan of \textit{nightlife spots}. However, when traveling with her family, she may prefer not to visit such locations. Hence, it is crucial to consider a user's context when recommending locations to her. It is also important to note that the user's context often introduces new constraints, not necessarily inclined to her opinion and interest. To this end, the main focus of the \textit{Text REtrieval Conference} (TREC) \textit{Contextual Suggestion} (TREC-CS) track\footnote{\url{https://sites.google.com/site/treccontext/}} in 2015~\cite{dean2015overview} and 2016~\cite{hashemi2016overview} was to improve location recommendation with the aid of contextual information. However, not many successful participants took into account context in their proposed approaches. Thus, applying contextual constraints still remains a challenge for context-aware POI recommendation.

\begin{figure}
    \centering
    \includegraphics[width=\textwidth]{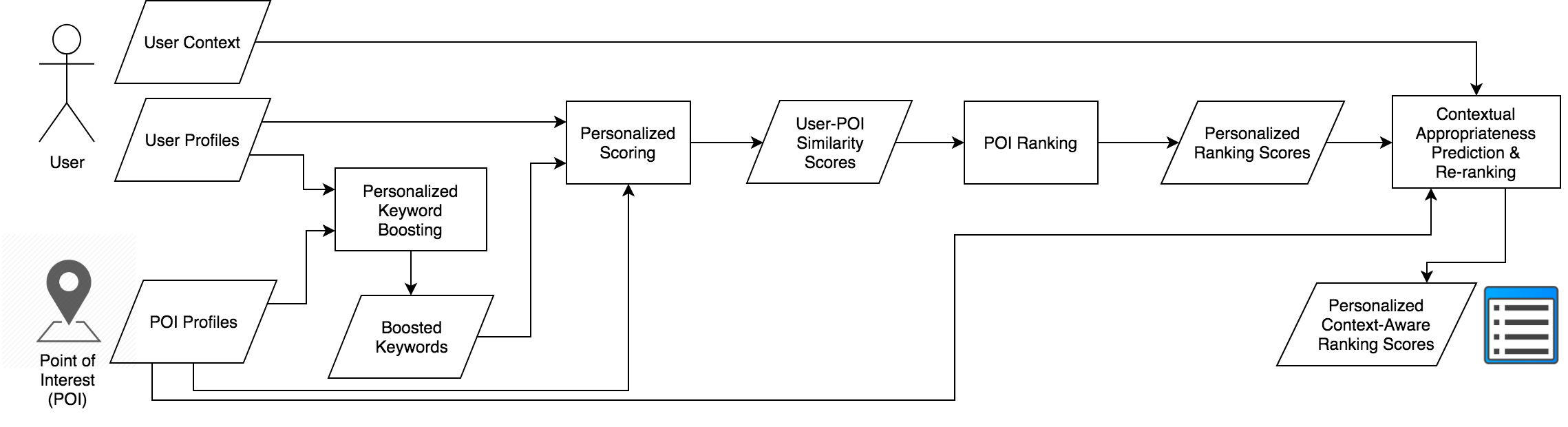}
    \caption{Overview of the proposed method.}
    \label{fig:overview}
\end{figure}

Given the easy access to the Internet and  availability of mobile devices such as cell phone, smart watches and tablets, users tend to leave their check-in data more often. However, writing a long review on such devices is not as trivial as is using a desktop computer. As a consequence, the majority of users rate locations without writing a review. Reviews contain critical information regarding a user's opinion and view about a location; for example a user's opinion about a location's view or staff. In order to compensate for the absence of such information, an LBSN can assist a user with a few related predefined \textit{tags}, from which the user can conveniently select those expressing her opinion. Predefined tags come very handy especially on smaller devices such as smart watches enabling users to express themselves with the aid of a couple of taps. Modeling users with such tags is very challenging since user tags are much more sparse compared to user ratings. Thus the traditional CF approach could not be applied for user tag modeling. 
Furthermore, in real-world POI recommendation scenarios for mobile devices, the top 10 locations are usually interesting to users~\cite{DBLP:journals/tist/ChengYKL16}, because of the screen size of a typical mobile device and limited effort a typical user spends to go through the recommendation list. Therefore, providing a personalized ranking to the user is crucial, making this task a \textit{top-k recommendation} task.

In this paper, we aim to answer the following research questions:	

\begin{itemize}
	\item RQ1: How can we model a user's interest and opinion based on her check-in record?
	\item RQ2: How can we leverage user tags to model users' interests and opinions more effectively to ultimately improve top-k POI recommendation effectiveness?
	\item RQ3: Could we use user tags to create a personalized model to reduce data dimensionality and address the data sparsity problem?
	\item RQ4: Could we model users to predict how they would tag a new location?
	\item RQ5: How can we incorporate users' context into top-k POI recommendation to improve the performance?
	\item RQ6: How can we integrate different aspects of information to generate personalized ranking considering both users' personal interests and contextual constraints?
\end{itemize}

In an effort to address these research questions, the contributions of this paper can be summarized as follows:
\begin{enumerate}
	 \item We introduce a set of relevance scores for measuring the similarity between a user's history and a location considering location's content and reviews.
	\item We present a probabilistic generative approach to find the mapping between location taste keywords and user tags thus modeling the personalized opinion of users about venues more accurately.
	\item We address the sparsity problem by performing personalized boosting of location keywords in a user's history.
	    \item We explore different machine learning models to predict user tags and evaluate the prediction effectiveness in terms of both tag prediction and recommendation effectiveness.    
	 \item We introduce a brand new dataset for predicting contextually appropriate locations and show how to predict the contextually appropriate locations given the user's current context and evaluate its effectiveness on recommendation.
    \item We evaluate several learning to rank techniques to incorporate boosting and tag prediction into our POI recommendation model using information from multiple LBSNs.
\end{enumerate} 

Figure \ref{fig:overview} shows an overview of our proposed method. In the first step, user and POI profiles are analyzed to perform personalized keyword boosting, resulting in a list of boosted keywords for each user (see Section \ref{sc:boosting}). Then the list of boosted keywords together with user and POI profiles are fed to the personalized scoring component to calculate the similarity scores between a given user and a POI. The scores are then passed to the ranking model (i.e., learning to rank) to produce a personalized ranked list of POIs for each user (see Section \ref{sc:suggestion}). Finally, given the user's current context, the level of contextual appropriateness of every venue is predicted (see Section \ref{sec-cxt}) and used to re-rank the personalized ranking scores, resulting in a personalized context-aware ranked list of POIs.

This paper extends over three years of past work on personalized context-aware suggestion that resulted in a number of published papers. In particular, in~\cite{DBLP:conf/airs/AliannejadiMC16} we first proposed a user model enrichment process leveraging user reviews from well-known LBSNs to support the venue suggestion process. In~\cite{alianSigir17} we tackled the POI appropriateness prediction problem, proposing a novel approach based on a user model enrichment process. In~\cite{DBLP:conf/ecir/Aliannejadi17} we combined multimodal information from multiple LBSNs to recommend POIs and studied how personalized mapping of location keywords to user tags could improve POI recommendation. Finally, in~\cite{alianSigir17collection} we made publicly available a contextual dataset that we created using crowdsourcing to incorporate the contextual constraints in our personalized recommender system.  All this work originated from our participation \cite{alian2015,alianTREC2016} to TREC-CS 2015 and TREC-CS 2016 where we consistently obtained the best run, demonstrating the effectiveness of our approaches under a common evaluation framework. This paper draws together content from all the above papers, presenting  a coherent synthesis of this full line of research. In addition, this paper extends our previous works as follows: 

\begin{itemize}
\item It introduces a unified recommendation framework, combining our previous works and showing how this can improve recommendation.
\item It presents deeper theoretical details of the proposed models, providing more discussion about keyword mapping, while elaborating on how to model user tag prediction as a sequence tagging problem.
\item It provides a more extensive experimentation on different datasets evaluating the proposed models from various perspectives. In particular, we evaluate our POI recommendation method in terms of personalized dimensionality reduction, user tag prediction, and contextual appropriateness prediction. We furthermore analyze the results pertaining each information component and reducing the number of visited POIs. 
\end{itemize}

This paper demonstrates that our proposed approach is able to outperform state-of-the-art strategies. In fact, experiments show that combining multimodal information from multiple LBSNs improves POI recommendation significantly. Moreover, we show that the proposed mapping of location keywords to user tags enables us to predict user tagging behavior effectively. We also show that predicting contextually appropriate locations and reranking suggestions according to their contextual appropriateness results in a more accurate top-k location recommendation.

The remainder of the paper is organized as follows. 
Before explaining our proposed recommendation approach, we describe two critical components of our model in Sections~\ref{sc:boosting} and \ref{sec-cxt}.
We first describe how we model the statistical mapping between user tags and venue taste keywords in Section~\ref{sec-mapping}. Then, we explore two directions to use this information. First, we explain how we use the computed mapping to reduce the dimensionality of venue taste keywords in Section~\ref{sec-boosting}. Second, we describe how we use the computed mapping as training data to learn the sequential tagging of user tags in Section~\ref{sec-user-tag}. The trained tagging models are then used to predict user tags for unseen venues.
Section~\ref{sec-cxt} describes our proposed contextual dataset and elaborates our proposed approach to predict contextual appropriateness of locations.
After describing the two main components of our model, in Section~\ref{sc:suggestion} we describe how we integrate them with other similarity measures in order to recommend POIs to users.
Section~\ref{Experiments} describes the evaluation protocol and Section~\ref{sec-results} presents our experimental evaluation, while Section~\ref{RelatedWork} reviews related work. We conclude our study in Section~\ref{Conclusions} proposing possible directions as future work.

\section{Personalized Keyword Boosting}\label{sc:boosting}
In this section we propose a probabilistic approach by which we map location keywords to user tags. Furthermore we propose two possible approaches of utilizing such additional information in order to enhance location recommendation. First, we propose to use the mapping as an additional information to reduce the dimensionality of location keyword space in order to address the data sparsity problem. Second, we use the mapping to train a sequence labeling model to predict user tags for a new location. We use the outcome of both directions to estimate the similarity of a location to a user in Section~\ref{sc:suggestion}.

\subsection{Personalized Keyword-Tag Mapping}\label{sec-mapping}
\begin{figure}[t]
    \centering
    \includegraphics[width=0.40\columnwidth]{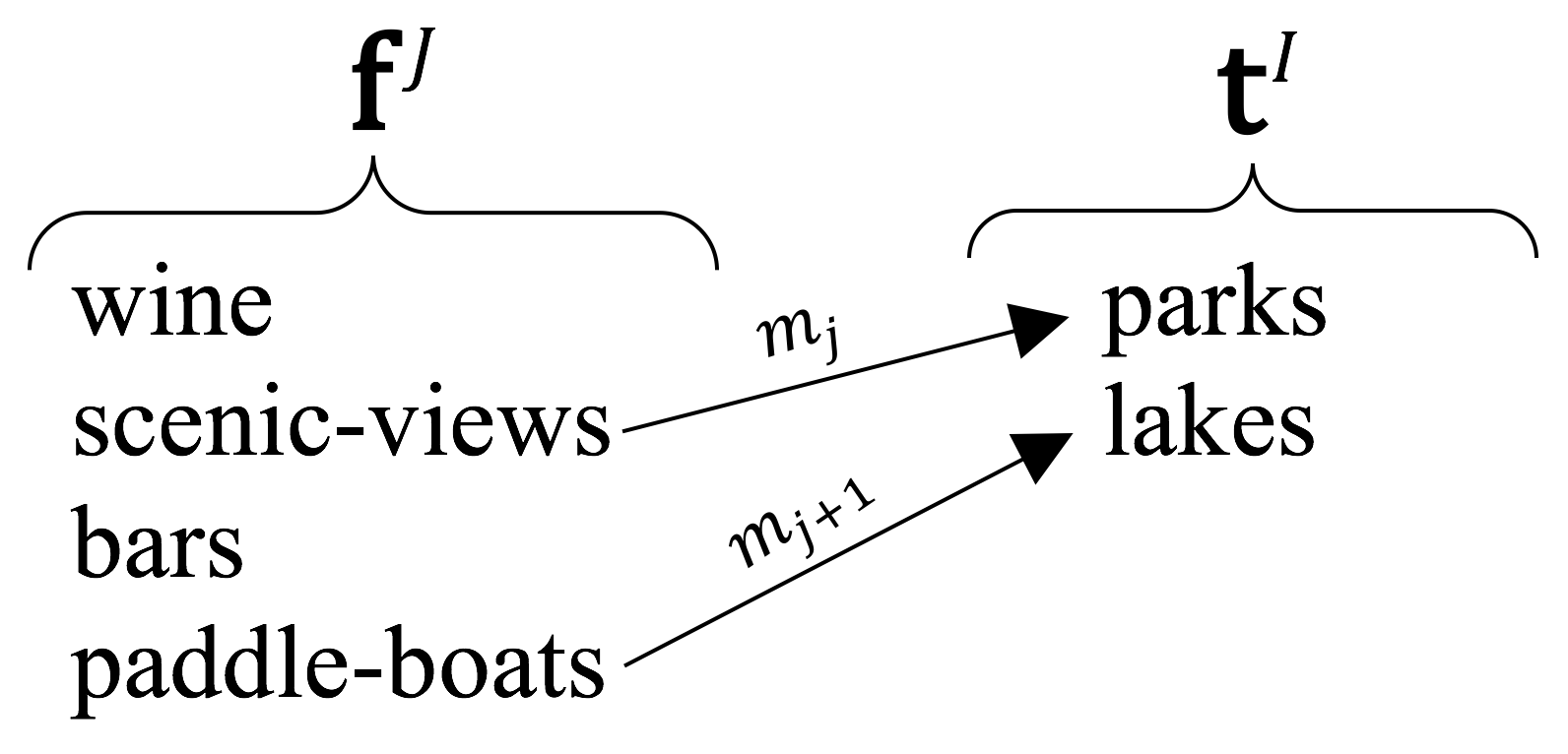}
    \caption{An example of mapping of $J=4$ location keywords to $I=2$ user tags.}
    \label{fg:mapping}
\end{figure}

In this section we present a probabilistic approach to map location keywords to user tags. We aim to find a meaningful correlation between the location content (e.g., keywords) and user tags since users annotate locations with tags based on both their personal views and locations' characteristics. We assume that the key characteristics that trigger the user's mind to annotate a location with a specific tag are of those listed in the location keywords. For example, tagging a location as \textit{healthy-food} is a result of user's personal view reflected in the location's characteristics (e.g., keywords). Hence, a user who believes \textit{vegan} foods are healthy may tag a \textit{vegan} location as \textit{healthy-food}, whereas another user with a different view may tag a \textit{sushi} place as \textit{healthy-food}. Therefore, user tags are dependent on both users views and locations' characteristics. That is why we need to find a meaningful mapping between user tags and location keywords to take into account locations' characteristics. The mapping needs to be personalized to model users' personal views.
Figure \ref{fg:mapping} depicts a real example mapping with a set of two user tags and four location keywords. Our ultimate goal is to determine the most likely mapping of location keywords to user tags, personalized for each user.

 For a given user $u$, let $\mathbf{f}^{J} = \langle f_1 \dots f_j \dots f_J \rangle$ be a sequence of location keywords. We aim to find the sequence of user tags $\mathbf{t}^{I} = \langle t_1 \dots t_i \dots t_I \rangle$. Note that $\mathbf{t}^{I}$ refers to a sequence named $\mathbf{t}$ with the length of $I$. Hence, $\mathbf{t}^{i}$ denotes a set with the size of $i$ ($\mathbf{t}^{i} = \langle t_1 \dots t_i \rangle$), whereas $t_{i}$ refers to the $i$-th item of a given sequence. 
Our aim is to find a user tag sequence maximizing $Pr(\mathbf{t}^{I}|\mathbf{f}^{J})$:

\begin{align}
    \label{eq:argmax}
    \begin{split}
        \hat{\mathbf{t}}^{I} & = \argmax_{\mathbf{t}^{I}}\{{Pr(\mathbf{t}^{I}|\mathbf{f}^{J})}\} 
                     = \argmax_{\mathbf{t}^{I}}\{{Pr(\mathbf{f}^{J}|\mathbf{t}^{I}) Pr(\mathbf{t}^{I})}\}~,
    \end{split}
\end{align}

 \noindent where $Pr(\mathbf{t}^{I})$ models user tags. In fact, given $\mathbf{t}^{I}$, this function determines to what extend $\mathbf{t}^{I}$ is likely to be generated by a specific user. It basically models the user's behavior of tag annotation regardless of location keywords. We fairly assume that users annotate locations with a specific tag independent of other tags. In other words, we assume zero-order dependence of user tags. Hence, we rewrite $Pr(\mathbf{t}^{I})$ as follows:

\begin{align}
    \begin{split}
        Pr(\mathbf{t}^{I}) & = p(I)\prod_{i=1}^{I} p(t_i|\mathbf{t}^{i-1}, I)
                  = p(I)\prod_{i=1}^{I} p(t_i|I)~,
    \end{split}
\end{align}
 \noindent where $\mathbf{t}^{i-1} = \langle t_1 \dots t_{i-1} \rangle$.

$Pr(\mathbf{f}^{J}|\mathbf{t}^{I})$ in \eqref{eq:argmax} models location keywords given a sequence of user tags $\mathbf{t}^{I}$. We need to find the optimum mapping between location keywords and user tags to optimally model location keywords given $\mathbf{t}^{I}$.
Therefore, we marginalize the probability $Pr(\mathbf{f}^{J}|\mathbf{t}^{I})$  over $\mathbf{m}^{J}$. We introduce $\mathbf{m}^{J}$ as the latent variable defining how location keywords are \textit{mapped} to user tags: $\mathbf{m}^{J} = \langle m_1 \dots m_j \dots m_J \rangle$, with $m_j \in \{1, \dots, I\} $:

\begin{align}
    \label{eq:margin}
    \begin{split}
    Pr(\mathbf{f}^{J}|\mathbf{t}^{I}) & = \sum_{\mathbf{m}^{J}}Pr(\mathbf{f}^{J}, \mathbf{m}^{J}|\mathbf{t}^{I})~,
    \end{split}
\end{align}
where

\begin{align}
    \label{eq:decomp}
    \begin{split}
                Pr(\mathbf{f}^{J}, \mathbf{m}^{J}| \mathbf{t}^{I}) = & ~p(\mathbf{m}^{J}|\mathbf{t}^{I}, I, J)p(\mathbf{f}^{J}|\mathbf{m}^{J}, \mathbf{t}^{I}, I, J) \\
                                      = & ~p(J|\mathbf{t}^{I})\prod_{j=1}^{J} [ p(m_j | \mathbf{m}^{j-1}, J, \mathbf{t}^{I}, I)p(f_j|\mathbf{f}^{j-1}, \mathbf{m}^{J}, J, \mathbf{t}^{I}, I) ]~,
    \end{split}
\end{align}
 \noindent where $\mathbf{m}^{i-1} = \langle m_1 \dots m_{i-1} \rangle$.
We also assume a zero-order dependence for both $m_j$'s and $f_j$'s. 
Note that given the limited amount of data and its sparsity, we make some assumptions in order to reduce the number of parameters.
Therefore, we consider $p(J|\mathbf{t}^{I})$ only depends on $J$ and $m_j$ is only dependent on the length of the user tag sequence $I$. We also assume that $f_j$ depends only on $t_{m_j}$, i.e., the user tag associated to $f_j$ according to the mapping. Notice that since $\mathbf{f}^{J}$ denotes the sequence of location keywords of size $J$ and $\mathbf{t}^{I}$ denotes the sequence of user tags of size $I$, therefore $Pr(\mathbf{f}^{J}, \mathbf{m}^{J}| \mathbf{t}^{I})$ also depends on the length of both sequences.
Consequently, \eqref{eq:decomp} is simplified as follows:
\begin{align}
    \begin{split}
        Pr(\mathbf{f}^{J}|\mathbf{t}^{I}) = p(J)\sum_{\mathbf{m}^{J}}\prod_{j=1}^{J}p(m_j|I)p(f_j|t_{m_j})~.
    \end{split}
    \label{eq:form}
\end{align}

\subsection{Parameter Estimation Based on Expectation-Maximization}\label{sec-em}
Assume that we have $N$ pairs of training samples as in $S = \{(\mathbf{f}_{(1)},\mathbf{t}_{(1)}), \dots, (\mathbf{f}_{(n)},\mathbf{t}_{(n)}), \dots, (\mathbf{f}_{(N)},\mathbf{t}_{(N)})\}$, the log-likelihood function for the training samples would be:

\begin{align}
    \label{eq:logll}
    \begin{split}
        F(\vartheta) =\sum_{n=1}^{N}\sum_{j=1}^{J_n}\log\sum_{i=0}^{I_n}p(i|I_n)p(f_{jn}|t_{is})~,
    \end{split}
\end{align}
where $\vartheta := \{p(i|I), p(f|t)\}$ are the free parameters. To solve the parameter estimation problem of \eqref{eq:logll}, we follow the Maximum Likelihood (ML) criterion subject to the constraint $\sum_f p(f|t)=1$, for each user tag $t$. 
We use Lagrange multipliers to make the optimization problem unconstrained. However, since we introduced hidden variables into our model (Equation \eqref{eq:margin}), there is no closed-form solution to this optimization problem. 
Therefore, we follow the iterative process of the \textit{Expectation-Maximization} (EM) algorithm.

To be able to follow the EM algorithm, we first define $Q(\vartheta, \hat{\vartheta})$ as:

\begin{align}
    \label{eq:q}
    \begin{split}
        Q(\vartheta, \hat{\vartheta}) & = Q\big(\{p(i|I), p(f|t)\};\{\hat{p}(i|I), \hat{p}(f|t)\}\big) \\
                                    & = \sum_{n=1}^{N}\sum_{j=1}^{J_n}\sum_{i=0}^{I_n}\gamma_n(i|j,J_n,I_n)\log\big\{\hat{p}(i|I_n)\hat{p}(f_{jn}|t_{in})\big\} \\
                                    & = \sum_{n=1}^{N}\sum_{j=1}^{J_n}\sum_{i=0}^{I_n}\frac{p(i|I_n)p(f_{js}|t_{is})}{\sum_{i'=0}^{I_n}p(i'|I_n)p(f_{js}|t_{i's})}\log\big\{\hat{p}(i|I_n)\hat{p} (f_{jn}|t_{in})\big\}~,
    \end{split}    
\end{align}
where $\gamma_n(i|j,J_n,I_n)$ is the posterior probability, defined as:
\[
    \gamma_n(i|j,J_n,I_n) = \frac{p(i|I_n)p(f_{js}|t_{is})}{\sum_{i'=0}^{I_n}p(i'|I_n)p(f_{js}|t_{i's})}~.
\]

According to the EM algorithm, we follow an iterative procedure for parameter estimation. After defining the relative objective function, $Q(\vartheta, \hat{\vartheta})$, we follow the usual steps of the algorithm:
\begin{enumerate}
    \item {E-step:} calculate $Q(\vartheta, \hat{\vartheta})$ for all training samples in $S$ with the previous estimate of $\vartheta$.
    \item {M-step:} optimize $Q(\vartheta, \hat{\vartheta})$ over $\hat{\vartheta}$.
\end{enumerate}

We start the algorithm with uniform values for the parameters and follow the EM steps until convergence.

In the following we describe two possible directions to use the computed mapping.

\subsection{Location Keywords Boosting}\label{sec-boosting}
\begin{figure}
    \includegraphics[width=0.6\textwidth]{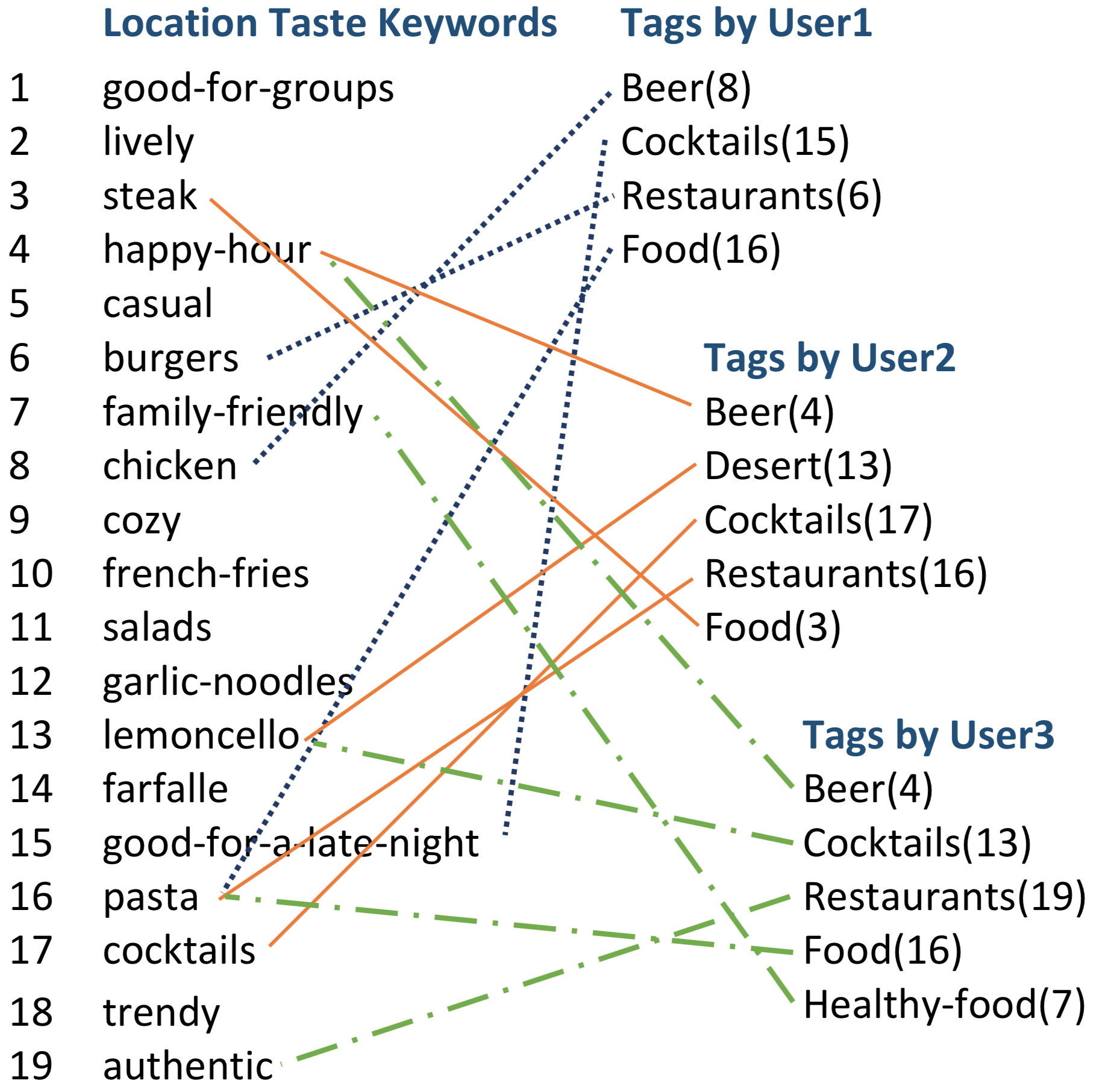}
    \caption{A sample of tags from three different users assigned to one single location and the calculated mapping. The lines connect each user tag to their mapped location keywords. Also, the index number of the mapped location keywords is written in parentheses for more convenient reading.}
    \label{fig-map-sample}
\end{figure}

After finding the optimum mapping between the user tags and location keywords, we aim to use this additional knowledge in our system to address the sparsity problem and eventually enhance the recommendation performance. Take Figure~\ref{fig-map-sample} as a real example of such mapping from our dataset. As we can see, 19 taste keywords from one location are illustrated together with tags for the same location by 3 different users . Not surprisingly, the 3 sets of tags have some in common such as ``beer'' and ``cocktails.'' However, each user has her own personal opinion and therefore her personal set of tags. The lines and the numbers in parentheses represent the result of our proposed mapping for these 3 users. Every mapped item is based on the user personal preference and behavior with respect to all locations in her history. As an example, we take one of the user tags that is common between the three users: ``cocktails.'' What is interesting about this tag is that each user maps it to a different location keyword. For User1 ``cocktails'' is mapped to ``good-for-a-late-night,'' for User2 to ``cocktails'' and for User3 to ``lemoncello.'' All three location keywords are good candidates to be mapped to ``cocktails'' user tag, however, as we argued \textit{each user has her own reasons to tag the same location with a different tag}. 

After the observation of Figure~\ref{fig-map-sample}, we assume that among the $I$ location keywords, we can determine $J$ keywords that are mapped to user tags and presumably are more interesting to the user. As in the example of Figure~\ref{fig-map-sample}, the number of location keywords ($I=19$) is much higher than user tags ($J_{user1}=4$). Therefore, by boosting the mapped location keywords in our model we achieve two main goals: 1) reduce the location keyword space dimensions drastically (e.g., $19 \rightarrow 4$) and; 2) use the valuable information given by the users to detect those  location keywords that are more interesting to each users. 

Formally, let $\mathbf{f}^I=\langle f_1 \dots f_I \rangle$ be the set of keywords of a location and $\mathbf{\hat{f}} \in \mathbf{f}^I$ be the set of location keywords which are mapped to user tags.
According to the result of our probabilistic mapping, we assume that there is a strong correlation between $\mathbf{\hat{f}}$ and the user's interest. In other words, the keywords in $\mathbf{\hat{f}}$ correlate more to the user's interest as opposed to the other ones in $\mathbf{f}^I$.  Hence, we boost $\mathbf{\hat{f}}$ to model the user's interests, reducing the data dimensionality from $I$ to $|\mathbf{\hat{f}}|$. This helps us to address the data sparsity problem. The personalized boosted location keywords are used for POI recommendation (see Section \ref{sc:suggestion}).

\subsection{User Tag Prediction}\label{sec-user-tag}
As an alternative approach, we explore three models to predict user tags. We utilize the result of our mapping model ($\mathbf{m}$) between location keywords and user tags to train a model being able to predict user tags for a new location. We predict user tags for an unseen location as an alternative to keyword boosting. We explore this direction for two reasons: 1) to see how we can predict a user behavior in terms of tag annotation and; 2) to compare the effect of user tag prediction against location keyword boosting to see which strategy is able to enhance the recommendation more effectively. We follow two approaches to predict user tags: 1) we use the maximum likelihood criterion with our estimated parameters to generate the most likely set of user tags given a set of location keywords and; 2) we model the user tag prediction as a sequence labeling problem enabling us to apply different sequence labeling models.

\partitle{Maximum Likelihood.}
Here we describe how we follow the Maximum Likelihood to leverage the learned mapping parameters in order to predict user tags for a new POI. As we mentioned in relation to \eqref{eq:argmax} in Section~\ref{sec-mapping}, given a set of location keywords from Foursquare, $\mathbf{f}^J$, we aim to compute the most probable set of user tags, $\hat{\mathbf{t}}^I$. Once the model parameters are estimated using EM (see Section~\ref{sec-em}), one approach to predict user tags given a set of location keywords is to follow the maximum likelihood criterion (see \eqref{eq:argmax}) to generate the most likely set of user tags. We search the space of user tag probabilities to find the optimum sequence of user tags following a Viterbi-like algorithm.

\partitle{Sequence Labeling.}
In the following, we begin with explaining how we model user tag prediction as a sequence labeling problem. Furthermore, we introduce the set of features we choose to train the tagging models as well as the tagging models we adopt.
Assuming we have $N$ sample mapped pairs of user tags and location keywords for each user: $S = \{(\mathbf{f}_{(1)},\mathbf{t}_{(1)}), \dots, (\mathbf{f}_{(n)},\mathbf{t}_{(n)}), \dots, (\mathbf{f}_{(N)},\mathbf{t}_{(N)})\}$ with $N$ corresponding mappings. That is, $M = \{\mathbf{m}_{(1)}, \dots, \mathbf{m}_{(n)}, \dots, \mathbf{m}_{(N)} \}$. We should model the tag prediction problem as a sequence labeling problem: given a sequence of location keywords we aim to predict the most probable sequence of user tags. In order to do this, we need to adapt the form of the training data.

As in a general sequence labeling problem, we need to assign a label from the target space to each item in the source space. Therefore, we should assign a label to all location keywords even if they are not mapped to any user tag. To this end, we automatically annotate location keywords following these steps:
\begin{enumerate}
    \item For each $f_i \in \mathbf{f}_{(n)}$ mapped to a user tag with $\mathbf{m}_{(n)}$, we annotate $f_i$ with its corresponding user tag $m_j$.
    \item For each $f_i \in \mathbf{f}_{(n)}$ not mapped to a user tag with $\mathbf{m}_{(n)}$, we annotate $f_i$ with ``null.''
\end{enumerate}

As for the example of Figure~\ref{fig-map-sample}, a sample training sequence would be as follows:

\begin{align*}
    \begin{split}
        \underbrace{ \mbox{burgers}}_{restaurants}, \underbrace{ \mbox{chicken}}_{beer}, \underbrace{ \mbox{good-for-a-late-night}}_{cocktails}, \underbrace{ \mbox{pasta}}_{food}, \underbrace{ \mbox{good-for-groups}}_{null}, \underbrace{ \mbox{lively}}_{null}, \underbrace{ \mbox{steak}}_{null}, \underbrace{ \mbox{happy-hour}}_{null}, \dots
    \end{split}
\end{align*}

As we can see, each of the location keywords is annotated with a tag. Therefore, it is straightforward to use this data as training samples for a sequence tagger.    
We adopt two models as taggers: \textit{Conditional Random Fields} (CRF)~\cite{DBLP:conf/icml/LaffertyMP01} and another tagger based on \textit{Support Vector Machines} (SVM)~\cite{DBLP:conf/acl/KudoM03}. The main advantage of these models is that they are discriminative. Discriminative tagging models have proven to be more effective for sequence labeling mainly because they normalize the model over the whole training set, resulting in a more generalized model. Furthermore, discriminative tagging models accept a wider range of features, something that is of great importance to some applications.

As features of the sequence labeling models, for a user tag at position $j$, $t_j$, we only consider the location keyword at the same position, $f_j$. That is to follow our assumption that user tags only depend on one location keyword and since we assume that user tags are independent, we use a zero-order tagger model. 

In this section, we first introduced a probabilistic framework for finding the mapping between location content (i.e., keywords) and user-annotated tags. Then, we described how this information can be leveraged to reduce the dimensionality of location keywords and hence to address the data sparsity problem. We also explored modeling this problem as a sequence labeling problem and used some state-of-the-art techniques to predict user tags given a new POI's keywords. In Section~\ref{sc:suggestion}, we show how we use the two mentioned directions to enhance POI recommendation.

\begin{table}[t]
	\caption{Description of different contextual information dimensions.}
	\label{tab-cxt-desc}
    \begin{tabular}{llll}
        \toprule
         Context  & Value & Short Reference & Description \\
         \midrule
         Duration & Day Trip & day-trip & The duration of the trip is one day. \\
                  & Night Out & night-trip & The duration of the trip is a night out. \\
                  & Weekend Trip & weekend-trip & The trip lasts for a weekend. \\
                  & Longer & longer-trip & The trip lasts longer than a weekend. \\
        \midrule
        Group     & Alone & alone & The person travels alone. \\
                  & Friends & with-friends & The person travels with her friends. \\
                  & Family & with-family & The person travels with her family. \\
                  & Other & with-others & The person travels with a group \\
                  & & & other than family and friends.\\
        \midrule
        Type      & Business & business-trip & The type of the trip is business. \\
                  & Holiday & holiday-trip & The type of the trip is holiday. \\
                  & Other & other-trip & The type of the trip is other than \\
                  & & & holiday and business, e.g., medical. \\
        \bottomrule
         
    \end{tabular}
\end{table}

\begin{table}[t]
    \caption{Examples of contextual features generated using crowdsourcing.}
    \begin{tabular}{llS[table-format=1.2]}
        \toprule
         Category & Context & {$F_{\text{app}}(cat,cxt)$} \\
         \midrule
         Beach & Trip type: Holiday & +1.0 \\
         College \& University & Trip duration: Weekend & -1.0 \\
         Shop \& Service & Trip type: Holiday & +0.71 \\
         Museum & Trip type: Business & -0.66 \\
         Pet Store & Trip duration: Weekend & -0.18 \\
         Medical Center & Trip type: Other & 0.0 \\         
         \bottomrule
    \end{tabular}
    \label{tab:features}
\end{table}

\section{Contextual Appropriateness Prediction}\label{sec-cxt}

In this section, we first define the problem of predicting the contextual relevance of locations. Then we present the set of features that we use to train the appropriateness classifier and introduce the dataset that we collected to train the classifier. The computed contextual relevance scores are then used to re-rank a ranked list of POIs in Section~\ref{sc:suggestion}.

Let $V = \{v_1, \dots, v_n\}$ be a set of locations and $C_x=\{cx_1,\dots,cx_m\}$ a set of contextual descriptors. Our aim is to predict whether it is appropriate for a user to visit a location $v_i \in V$ under a given context $C_x$. Different contextual dimensions define user's preferences, constraints, or requirements and are listed as follows: \textbf{Trip type} (holiday, business, other), \textbf{Trip duration} (day trip, night out, weekend trip, longer) and \textbf{Group type} (alone, family, friends, other). More information could be found in \autoref{tab-cxt-desc}. We model the problem as binary classification considering location categories and contextual descriptors as classification features.

\subsection{Contextual Features}\label{sec:features}
In this section, we describe the features that we used to train the appropriateness classifier.
The degrees of appropriateness between location categories and contextual descriptors constitute our features. 

We define a contextual feature function as follows:

\begin{definition}
    \label{df:contextfeature}
    A \textbf{contextual feature}, $F_{\text{app}}(cat, cxt)$, is a function determining the relevance of a POI category, $cat$, to a contextual dimension, $cxt$. $F_{\text{app}}(cat, cxt)$ ranges between $-1$ and $+1$ with $-1$ representing absolute inappropriateness and $+1$ absolute appropriateness.
\end{definition}

For instance, assume a user wants to visit a location with category \textit{nightlife-spot}, and her context is described as follows: holiday-trip, with-family, weekend-trip. The three features of this example are the appropriateness value between the location category and each of the contextual dimensions. 
Therefore, the features are $F_{\text{app}}(\text{nightlife-spot}, \text{holiday-trip})$, $F_{\text{app}}(\text{nightlife-spot}, \text{with-family})$, and $F_{\text{app}}(\text{nightlife-spot}, \text{weekend-trip})$. 

In many cases, determining such contextual features is intuitive and can be done by one human annotator. However, there are several features that human annotators cannot agree on, like for example $F_{\text{app}}(\text{office}, \text{with-friends})$, 
$F_{\text{app}}(\text{food-and-drink-shop}, \text{business-trip})$, or $F_{\text{app}}(\text{stadium},$ $\text{night-out-trip})$. Hence, we define two classes of features: objective and subjective. \textit{Objective} features are those that the annotators quickly agree on. This suggests that a user would be very likely to agree with the annotators on the objective features. Therefore, we can conclude that objective features potentially influence a user's decision of visiting a location. As in the previous example, supposedly, everyone would consider going to a nightlife spot with family is \textit{not} appropriate. Thus a user who regularly goes to nightlife spots might change her mind when she is traveling with her family. As we saw in this example, such \textit{objective} features can directly change users' decisions adding contextual constraints to the model. \textit{Subjective} features, in contrast, have less impact as they mainly depend on the user's opinion and personal preferences. If the annotators did not agree on a feature, we would not be able to predict a user's opinion. Therefore, we cannot predict the influence of subjective features on a user's decision. 

We determined the level of subjectivity or objectivity of features via a crowdsourcing task.
In the task, we asked the workers to judge if a location category is appropriate for a context descriptor (e.g., $cat=\text{nightlife-spot}$ and $cxt=\text{with-family}$). We asked at least five different assessors to judge each category-context pair. In the context of this paper, we define those pairs with high agreement rate between the workers as \textit{objective}, while we consider those lacking assessors agreement as \textit{subjective}.
More details on how we created the dataset can be found in our previous work~\cite{alianSigir17collection}.

Table~\ref{tab:features} lists some example features from our dataset. As we can see in this table, lower values for $|F_{\text{app}}(cat,cxt)|$ mean that the features are more subjective.
We created the contextual features for all pairs of 11 contextual dimensions and the 177 most frequent categories of Foursquare category tree\footnote{\url{https://developer.foursquare.com/categorytree}}. 
Overall we generated 1,947 contextual features from 11,487 judgments\footnote{The dataset if freely available on request.}.

\subsection{Training the Classifier}
As described earlier, we formulate contextual appropriateness as binary classification. In Section~\ref{sec:features} we explained how we created the contextual features. Here, we describe another dataset for training the appropriateness classifier using our features. We randomly selected 10\% of the data from TREC-CS 2016 dataset. We created another crowdsourcing task for annotating the training data. We asked workers to assess if it is appropriate that a user with a full description of context (e.g., Holiday, Friends, Weekend) to visit a location category (e.g., Bar). Each instance in the dataset is considered as appropriate only if at least two of the three assessors voted for their appropriateness. We train the contextual appropriateness classifier on 10\% of the data from TREC-CS 2016 to predict the remaining 90\% of TREC-CS 2016 and the whole TREC-CS 2015 dataset. 
We applied a wide range of classifiers for this task. However, we only report the best results that were obtained with SVM~\cite{cortes1995support}. The predicted contextual relevance score is denoted as $S_{cxt}$ and we use it to re-rank the personalized location ranking (see Section~\ref{sc:suggestion}).

\section{ Recommendation based on Information from Multiple Location-Based Social Networks}\label{sc:suggestion}
After explaining the two major components of our proposed approach in Sections~\ref{sc:boosting} and \ref{sec-cxt}, here, we explain our way of performing POI recommendation, exploiting the scores from multiple LBSNs. We describe two sets of scores: the frequency-based and review-based scores. 
We use the frequency-based score to incorporate the boosted keywords (Section~\ref{sec-boosting}), the predicted user tags (Section~\ref{sec-user-tag}), as well as other types of information.
We also demonstrate how we combine different scores to produce the personalized location ranking using learning to rank.

\subsection{Frequency-based Score}\label{ContextModeling}
We base the frequency-based scores on the assumption that users prefer the type of locations that they like more frequently and rate them positively\footnote{We consider reviews with rating [4, 5] as positive, 3 as neutral, and [1, 2] as negative.}. Therefore, we create positive and negative profiles considering the content of locations in the user's check-in history and calculate the normalized frequencies as they appear in her profile. Then we compute a similarity score between the user's profile and a new location. For simplicity, we only explain how to calculate the frequency-based score using location categories. The method can be easily generalized to calculate the score for other types of information.

Let $u$ be a user and $h_u = \{v_1, \dots, v_n\}$ her history of check-ins. Each location has a list of categories
$C(v_i) = \{c_1, \dots, c_k\}$. We define the user category profile as follows:
\begin{definition}
    \label{df:profile}
    A \textbf{Positive-Category Profile} is the set of all unique categories belonging to locations that user $u$ has previously rated positively. A \textbf{Negative-Category Profile} is defined analogously for locations that are rated negatively.
\end{definition}
Each category in the positive/negative category profile is assigned with a user-level normalized frequency. We define the user-level normalized frequency for a category as follows:
\begin{definition}
    \label{df:freq}
    A \textbf{User-level Normalized Frequency} for an item (e.g., category) in a profile (e.g., positive-category profile) for user $u$
    is defined as: 
    
    \[
       \mathrm{cf}_u^+(c_i) = \frac{\sum_{v_k \in h_u^+}\sum_{c_j \in C(v_k), c_j=c_i} 1}{\sum_{v_k \in h_u}\sum_{c_j \in C(v_k)} 1}~,
    \]
     where $h_u^+$ is the set of locations that $u$ rated positively. We calculate user-level normalized frequency for negative categories, $\mathrm{cf}_u^-$, analogously. 
    
\end{definition}

We create positive/negative category profiles for each user based on Definitions \ref{df:profile} and \ref{df:freq} . Given a user $u$ and candidate location $v$, the frequency-based similarity score based on location categories, $S_{cat}(u,v)$, is calculated as follows:
\begin{equation}
    \label{eq:cat}
    S_{cat}(u,v) = \sum_{c_i \in C(v)}\text{cf}_u^+(c_i) - \text{cf}_u^-(c_i)~.
\end{equation}

We follow the same procedure to calculate a frequency-based score based on other types of information as listed below:
\begin{itemize}
	\item \textbf{$S_{key}$:} We consider location taste keywords (instead of categories) to compute the similarity score between a given user's profile and a candidate location.
	
	 \item \textbf{$S_{boost}$:} As for boosting, for each user we follow Definition \ref{df:profile} to create positive and negative \textit{boosted location taste keyword profiles}. We consider only the location taste keywords that are mapped to user tags (see Section~\ref{sec-boosting}). Given a candidate location, we calculate boosted keywords similarity score according to Definition \ref{df:freq} and \eqref{eq:cat}.
		
	\item \textbf{$S_{ml}$:}
	Following Definitions \ref{df:profile} and \ref{df:freq}, we create positive and negative \textit{user tag profiles} for each user.  However, since a candidate location does not naturally have user-assigned tags, we use our ML-based approach to predict user tags. The predicted user tags are then compared with the user's profile following \eqref{eq:cat} to calculate $S_{ml}$ similarity score.
	
	\item \textbf{$S_{crf}$:}
	We calculate $S_{crf}$ score similar to $S_{ml}$. We predict user tags for a candidate location using the trained CRF model. Then we follow \eqref{eq:cat} to calculate the $S_{crf}$ score comparing predicted user tags with user profile.
	
	\item \textbf{$S_{svm}$:}
	This score is also calculated like $S_{ml}$. For a candidate location, we predict user tags using our trained SVM-based tagging model.  The predicted user tags are then compared with the user's profile using \eqref{eq:cat} resulting in $S_{svm}$.
\end{itemize}

\subsection{Review-Based Score}\label{sec:rev}
Modeling a user only on locations' content is general and does not determine why the user enjoyed or disliked a POI.
The content of locations is often used to infer ``which type'' of POIs a user likes. On the other hand, reviews express the reasons of users' ratings~\cite{yang2015opinions}. Since there could be a lack of explicit reviews from the user, we tackle this sparsity problem using reviews of other users who gave a similar rating to the location. 
We follow the same idea of Yang et al.~\cite{yang2015opinions}, that is, a user's opinion regarding a location could be learned based on the opinions of other users who rated the same location similarly.

We calculate the review-based score using a binary classifier. We model this problem as binary classification since a user, before visiting a new city or location, would get a positive or negative impression of the location after reading the online reviews of other users. We assume that a user compares the characteristics of a location and the opinions which are expressed by other users in their reviews to her expectations and interests.
A user would be convinced of visiting a particular location if the reviews satisfy her expectations up to a certain point.
An alternative to binary classification would be a regression model. However, we assume that users behave similarly to a binary classifier when they read online reviews before deciding on whether to visit a venue or not. For example, assume a user reads a few positive and negative online reviews about a POI and measures how similar the mentioned qualities are to her expectations. Finally, depending on the balance between the positive remarks and the negative ones, she makes a binary decision (i.e., whether to go or not). We see this behavioral pattern similar to that of a binary classifier: it learns from the positive and negative samples and compares the learned parameters with a test sample and assigns its label accordingly.
Furthermore, due to data sparsity, grouping ratings as positive and negative aids us to model users more effectively.

We train binary classifier using the reviews from the locations in a user's check-in history. 
The positive training samples for user $u$ are positive reviews of locations that were liked by $u$. Likewise, the negative reviews of locations that $u$ disliked constitute the negative training samples.
We decided to ignore the negative reviews of liked locations and positive reviews of disliked locations since they are not supposed to contain any useful information.

We consider TF-IDF score of terms in reviews as features. We trained many classifiers but SVM outperformed all other models. Therefore, we choose SVM and consider the value of its decision function as the review-based score and refer to it as $S_{rev}$. The decision function gives us an idea on how relevant a location is to a user profile. 

\subsection{Location Ranking}\label{sec:ranking}
After defining the mentioned relevance scores, here we explain how we combine them.
Given a user and a list of candidate locations, we calculate the mentioned scores for each location and combine them to create a ranked list of locations. We adopt several learning to rank\footnote{We use RankLib implementation of learning to rank: \url{https://sourceforge.net/p/lemur/wiki/RankLib/}} techniques to rank the candidate locations since they have proven to be effective for similar tasks~\cite{DBLP:journals/ftir/Liu09,yang2015opinions}. In particular, we examine the following learning to rank techniques: AdaRank, CoordinateAscent, RankBoost, MART, LambdaMART, RandomForest, RankNet, and ListNet. We introduce four models using different combinations of the scores as mentioned in Table~\ref{tab:models}.

\begin{table}[]
    \centering
    \caption{Four proposed models using different combination of similarity scores.}
    \begin{tabular}{lllll}
    \toprule
               &  Category & Review    & Keywords            & Context\\
    \midrule               
         UT-ML & $S_{cat}$ & $S_{rev}$ & $S_{key}$, $S_{ml}$ & $S_{cxt}$ \\
         UT-CRF & $S_{cat}$ & $S_{rev}$ & $S_{key}$, $S_{crf}$ & $S_{cxt}$ \\
         UT-SVM & $S_{cat}$ & $S_{rev}$ & $S_{key}$, $S_{svm}$ & $S_{cxt}$ \\
         PK-Boosting & $S_{cat}$ & $S_{rev}$ & $S_{key}$, $S_{boost}$ & $S_{cxt}$ \\
         \bottomrule
    \end{tabular}
    \label{tab:models}
\end{table}

\section{Experimental Settings}\label{Experiments}
In this section, we present the experimental settings including the datasets we used, compared methods, and evaluation process.

\subsection{Datasets} 
\partitle{Recommendation Effectiveness.}
We evaluate our approach on two benchmark collections, published by TREC. The collections are those used in the \textit{Batch Experiments}/\textit{Phase 2} of the TREC-CS track 2015~\cite{dean2015overview} and 2016~\cite{hashemi2016overview}. The task was to rank a list of candidate locations in a new city for a user, given her history of check-ins in other cities. The datasets were collected using crowdsourcing where each user rated 30 to 60 locations in one or two cities. In addition, each user may have tagged locations to explain why she likes them (i.e., user tags). Later, the same users were called to rate new POIs in another city as well as the contextual factors of their trip.
To get more information about locations, we crawled Yelp and Foursquare for reviews, categories, and taste keywords\footnote{The dataset is available on request.}. More specifically, for each location in the dataset, we formed a query from the location's name and city to find the corresponding Yelp and Foursquare profiles. To avoid irrelevant information, we verified the title and location of each result. Yelp is crawled mainly for reviews, whereas Foursquare mainly for location taste keywords. It is worth noting that we were able to find additional information for most of the locations from both LBSNs. Table \ref{tb:ds-stats} lists the key attributes of our crawled dataset. More details on the crawling process and structure of our dataset can be found in our previous work~\cite{alianSigir17collection}.
    
    \begin{table}
        \centering
        \caption{Statistical details of datasets}
        \begin{tabular}{l@{\quad}l@{\quad}l}
            \toprule
            & TREC-CS 2015 & TREC-CS 2016 \\
            \midrule
             Number of requests & 211 & 442 \\
             Number of requests evaluated by TREC & 211 & 58 \\
             Number of locations & 8,794 & 18,808\\
             Number of locations crawled from Yelp & 6,290 & 13,604\\
             Number of locations crawled from Foursquare & 5,534 & 13,704\\
             Average reviews per location & 117.34 & 66.82\\
             Average categories per location & 1.63 & 1.57\\
             Average taste keywords per location & 8.73 & 7.89\\
             Average user tags per user & 1.46 & 3.61\\
             Number of distinct user tags & 186 & 150\\
             \bottomrule
        \end{tabular}
        \label{tb:ds-stats}
    \end{table}

\partitle{Dimensionality Reduction.}
We evaluate the dimensionality reduction effectiveness on the same datasets as we do for recommendation effectiveness. We compare the performance of our proposed model to a well-known dimensionality reduction model in terms of recommendation effectiveness. Therefore, we use the same datasets used to evaluate recommendation effectiveness, however, we provide more details and discussion related to dimensionality reduction.

\partitle{User Tag Prediction.}
Since the test set in TREC-CS 2015 and TREC-CS 2016 do not include user tags for locations, we need to evaluate the user tag prediction on the training datasets. Therefore, we randomly split the TREC-CS 2015 and TREC-CS 2016 training sets into: training, development, and test set. We train the taggers using the new training set, tune them using the new development set and evaluate them with the new test set. As part of the evaluating recommendation effectiveness we show how different taggers can improve the recommendation; however, the aim of this experiment is to show how accurately we can model the user interests and tagging behavior. The statistical details of the tags dataset is listed in Table~\ref{tab-db-tag}.

\begin{table}
	\centering
	\caption{Statistical details of user tagging dataset}
	\begin{tabular}{l@{\quad}l@{\quad}l@{\quad}l}
	\toprule
	& Training Set & Test Set \\ 
	\midrule
	Number of instances & 20,148 & 5,037 \\
	Number of non-null tags & 102,667 & 25,541\\
	Number of null tags  & 54,444 & 13,954 \\
	Number of unique user tags & 156  & 121 \\
	Number of unique location keywords & 2,676 & 1,398 \\
	Average user tags per location &1.85 & 1.82 \\
	Average keywords per location & 5.10 & 5.07 \\
	\bottomrule
	\end{tabular}
	\label{tab-db-tag}
\end{table}

\subsection{Compared Methods}
\partitle{Recommendation Effectiveness.} We consider the best performing system in TREC-CS 2015 as our baseline. Moreover, we compare our proposed method with state-of-the-art context-aware POI recommendation methods. We also compare our proposed PK-Boosting with other models based on user tag prediction (i.e., UT-ML, UT-CRF, and UT-SVM).

\begin{itemize}
	\item \textit{LinearCatRev} is our previous work~\cite{DBLP:conf/airs/AliannejadiMC16} which is the best performing model of TREC-CS 2015. It extracts information from different LBSNs and uses it to calculate category-based and review-based scores. Then, it combines the scores using linear interpolation. We choose this baseline for two reasons, firstly because it is the best performing system of TREC-CS 2015, and secondly because it also uses scores derived from different LBSNs.

	\item \textit{GeoSoCa} exploits geographical, social, and categorical correlations for POI recommendation~\cite{DBLP:conf/sigir/ZhangC15}. GeoSoCa models the geographical correlation using a kernel estimation method with an adaptive bandwidth determining a personalized check-in distribution. It models the categorical correlation by applying the bias of a user on a POI category to weigh the popularity of a POI in the corresponding category modeling the weighted popularity as a power-law distribution. We used the implementation of GeoSoCa released in~\cite{DBLP:journals/pvldb/LiuPCY17}. We did not include the social correlation component since such information does not exits in the datasets.

    \item \textit{n-Dimensional Tensor Factorization (nDTF)}~\cite{DBLP:conf/recsys/KaratzoglouABO10} generalizes matrix factorization to allow for integrating multiple contextual features into the model. We used the publicly available implementation of nDTF\footnote{\url{https://github.com/VincentLiu3/TF}}. Regarding the features, we include two types of features: (1) location based: category, keywords, average rating on Yelp, and number of ratings on Yelp (as an indicator of its popularity); (2) user based: age group and gender.

	\item \textit{UT-ML} differs from PK-Boosting in one score. For UT-ML, instead of the keyword boosting score ($S_{boost}$), we use the score based on the predicted user tags following maximum likelihood criterion. As we described in Section~\ref{sec-user-tag}, we also explored three different models as alternatives to PK-Boosting. Our aim is to study the impact of predicting user tags on recommendation effectiveness, compared to PK-Boosting. The other two alternative approaches are listed as follows.

	\item \textit{UT-CRF} predicts user tags using a trained CRF model. Then, for each venue-user pair, it computes the similarity between the predicted user tags and the user profile. Finally, it replaces the boosting score with the computed similarity score (see Section~\ref{sec-user-tag}). We used CRFSuite\footnote{\url{http://www.chokkan.org/software/crfsuite/}} implementation of CRF.

	\item \textit{UT-SVM} predicts user tags given a user-venue pair using an SVM-based tagging model. The boosting score is replaced by the similarity score between the user profile and predicted user tags (see Section~\ref{sec-user-tag}). We used YamCha\footnote{\url{http://chasen.org/~taku/software/yamcha/}} implementation of the SVM-based tagging model.

\end{itemize}

\partitle{Dimensionality Reduction.} 
In order to evaluate the keyword boosting approach from the perspective of dimensionality reduction, we also apply the following well-known dimensionality reduction method to reduce the location keywords dimension. In particular, we use \textit{PK-PCA.} PK-PCA uses \textit{Principal Component Analysis} (PCA) to reduce the dimensionality of location keywords. The corresponding score of keyword boosting is replaced by the score computed based on PCA.

\partitle{User Tag Prediction.}
As user tags contain very crucial information explicitly described by users, we aim to evaluate the effectiveness of user tag models. In particular, we evaluate the following models:
\begin{itemize}
\item \textit{Conditional Random Fields (CRF) Tagger}~\cite{DBLP:conf/icml/LaffertyMP01} models the sequence tagging problem in a discriminative manner. The tagger is based on binary features that are extracted from the text and optimized for the training data. 

\item \textit{SVM-based Tagger}~\cite{DBLP:conf/acl/KudoM03} is also a discriminative approach that trains one SVM classifier per tag. The model is an ensemble of all SVM classifiers.

\end{itemize}

\subsection{Evaluation Metrics}
We evaluate the recommendation effectiveness as well as the dimensionality reduction for the top-$k$ recommendation. We also evaluate the effectiveness of user tag prediction methods.

\partitle{Recommendation Effectiveness.}
In both TREC-CS 2015 and TREC-CS 2016 datasets, for each user $u$, the data, $S(u)$, is split into two sets: a number of locations visited in one or two cities constitute the training set  and a number of locations in a new city constitute the test set. Given a user $u$, if a recommended location in the test set is marked by the user as \emph{relevant}, it is a ``hit,'' otherwise it is a ``miss.''  
To perform a fair comparison, we choose the official evaluation protocol and metrics of TREC-CS for this task, which are P@5 (Precision at 5), nDCG@5 (Normalized Discounted Cumulative Gain at 5), and MRR (Mean Reciprocal Rank). Since the main focus in this task was to improve the location rankings, such evaluation metrics serve as perfect metrics. 

The relevance assessments for test sets are slightly different in TREC-CS 2015 and TREC-CS 2016. In TREC-CS 2015 relevance of a location to a user is defined with a binary value with 0 as irrelevant and 1 as relevant, whereas in TREC-CS 2016 users rated locations in the range of $-$2 to +2.
This also explains why the main evaluation metric in TREC-CS 2015 is P@5 as opposed to nDCG@5 in TREC-CS 2016. Both P@$k$ and nDCG@$k$ metrics are evaluated over $k$ top locations on the ranked list. Let $U$ be the set of users and $r_u^p$ be the rating score assigned by user $u$ to the location at the $i$\textsuperscript{th} rank of the list. Precision and nDCG values are calculated at the $k$\textsuperscript{th} position as follows:

\[
	P_u@k = \frac{\#hits_{u}@k}{k}~,
\]
\[
	nDCG_{u}@k = Z_u \sum_{i=1}^{k}\frac{2^{r_u^i}-1}{\log(1+i)}~,
\]

\noindent 
where $u$ is the given user, $Z_u$ is a normalization factor and $\#hits_{u}@k$ is the number of relevant locations for user $u$ in the top-$k$ locations of the ranked list. nDCG@$k$ and P@$k$ are the mean of nDCG$_{u}@k$ and P$_u@k$ over $U$ respectively. MRR is also calculated as follows:
\[
   MRR = \frac{1}{|U|}\sum_{u=1}^{|U|}\frac{1}{rank_u}~,
\]
 
 \noindent 
where $rank_u$ is the ranking of the first relevant location for user $u$. We conduct a 5-fold cross validation on the training data to tune our model. We determine the statistically significant differences using the two-tailed paired t-test at a $95\%$ confidence interval ($p < 0.05$). 

\partitle{Dimensionality Reduction.}
Since there is no ground truth data to evaluate dimensionality reduction methods, we evaluate the recommendation effectiveness using different dimensionality reduction methods to see how they enhance the overall recommendation. Therefore, we use the same evaluation metrics that we used for evaluating recommendation effectiveness.

\partitle{User Tag Prediction.}
Since we modeled the user tag prediction problem as a sequence-labeling problem, we evaluate the effectiveness of user tag prediction using the same metrics used for evaluating typical sequence-labeling problems such as \textit{Part of Speech} (POS) tagging. Therefore, we report Precision, Recall, and F-Measure for this experiment. Let $T_p$ the number of true positive and $F_p$ the number false positive non-null predicted tags. Then precision is calculated as follows:
\[
	Precision = \frac{T_p}{T_p + F_p}~.
\]
Given the number of false negatives, $F_n$, we also calculated recall as follows:
\[
	Recall = \frac{T_p}{T_p + F_n}~.
\]
F-measure is then defined as follows:
\[
	F-Measure = 2\times\frac{Precision \times Recall}{Precision + Recall}~.
\]

 \section{Experimental Results}\label{sec-results}
In this section, we first present the results for recommendation effectiveness. We also show the results for location keyword dimensionality reduction and user tag prediction. Furthermore, we study the effect of different sources and scores on recommendation effectiveness.

\subsection{Recommendation Effectiveness} 
Tables \ref{tb:preliminarily_results_2015} and \ref{tb:preliminarily_results_2016} demonstrate the performance of our approach compared with other methods for the TREC-CS 2015 and TREC-CS 2016 datasets, respectively. We choose the best performing learning to rank technique for each model we adopt the best performing learning to rank technique according to Tables \ref{tb:ltr2015} and \ref{tb:ltr2016}. It is worth noting that the best learning to rank technique for PK-Boosting is ListNet~\cite{listNet}. Tables \ref{tb:preliminarily_results_2015} and \ref{tb:preliminarily_results_2016} show that PK-Boosting outperforms the competitors in terms of the three evaluation metrics. This indicates that the proposed  PK-Boosting approach improves the performance of POI recommendation. This happens because the proposed approach for boosting location keywords addresses the data sparsity problem, while at the same time it captures user preferences more accurately. In contrast, the models UT-ML, UT-CRF, and UT-SVM introduce a prediction error, when predicting user tags for a candidate location. This error is then propagated to location ranking and subsequently degrades the models' performances. As we can see, GeoSoCa and nDTF exhibit the worst performance among all compared methods. This happens mainly because these methods rely on user-POI check-in associations among the training and test sets. In other words, there should be enough common POIs appearing in both the training and test sets, otherwise they fail to recommend unseen POIs. Hence, they suffer from the high level of sparsity on these datasets. In particular, the intersection of POIs in the training and test sets is 771 (out of 8,794) and 4 (out of 18,808) in TREC-CS 2015 and 2016, respectively. 

To compute the review-based classifier, we used various classifiers such as Na{\"i}ve Bayes and k-NN; however, the SVM classifier exhibited a better performance by a large margin. The SVM classifier is a better fit for this problem since it is more suitable for text classification, which is a linear problem with weighted high dimensional feature vectors. Also, we observed a significant difference between the number of positive reviews and negative reviews per location. Generally, locations receive more positive reviews than negative reviews and, in our case, this results in a unbalanced training set. Most of the classification algorithms fail to deal with the problem of unbalanced data. This is mainly due to the fact that those classifiers try to minimize an overall error rate. Therefore, given an unbalanced training set, the classifier is usually trained in favor of the dominant class to minimize the overall error rate. However, SVM does not suffer from this, since it does not try to directly minimize the error rate but instead tries to separate the two classes using a hyperplane maximizing the margin. This makes SVM more intolerant of the relative size of each class. Another advantage of linear SVM is that the execution time is very low and there are very few parameters to tune. 

\begin{table}
\centering
\caption{Performance evaluation on TREC-CS 2015. Bold values denote the best scores and the superscript * denotes significant differences compared to LinearCatRev. $\Delta$ values ($\%$) express the relative improvement, compared to LinearCatRev. For each model we report the scores using the best learning to rank technique (Table \ref{tb:ltr2015}).}
\label{tb:preliminarily_results_2015}
\begin{tabular}{l@{\quad}l@{\quad}r@{\quad}l@{\quad}r@{\quad}l@{\quad}r}
\toprule
 & \multicolumn{1}{c}{P@5} & \multicolumn{1}{c}{$\Delta(\%)$} & \multicolumn{1}{c}{nDCG@5} & \multicolumn{1}{c}{$\Delta(\%)$} & \multicolumn{1}{c}{MRR} & \multicolumn{1}{c}{$\Delta(\%)$} \\
\midrule
LinearCatRev & 0.5858  & \multicolumn{1}{c}{{-}} & 0.6055 & \multicolumn{1}{c}{{-}} & 0.7404 & \multicolumn{1}{c}{{-}} \\
GeoSoCa & 0.5147* & $-$12.14 & 0.5404* & $-$10.75 & 0.6918* & $-$6.56\\
nDTF & 0.5232* & $-$10.96 & 0.5351* & $-$11.63 & 0.6707* & $-$9.41 \\
UT-ML & 0.6224* & 6.25 & 0.6320* & 4.38 & 0.7496 & 1.24 \\
UT-CRF & 0.6249* & 6.67 & 0.6285 & 3.80 & 0.7434 & 0.41 \\
UT-SVM & 0.6219* & 6.16 & 0.6339* & 4.69 & 0.7553 & 2.01 \\
PK-Boosting & \textbf{0.6259}* & \textbf{6.85} & \textbf{0.6409}* & \textbf{5.85} & \textbf{0.7704}* & \textbf{4.05} \\
\bottomrule
\end{tabular}

\bigskip

\centering
\caption{Performance evaluation on TREC-CS 2016. Bold values denote the best scores and the superscript * denotes significant differences compared to LinearCatRev. For each model we report the scores using the best learning to rank technique (Table \ref{tb:ltr2016}).}
\label{tb:preliminarily_results_2016}
\begin{tabular}{l@{\quad}l@{\quad}r@{\quad}l@{\quad}r@{\quad}l@{\quad}r}
\toprule
 & \multicolumn{1}{c}{P@5} & \multicolumn{1}{c}{$\Delta(\%)$} & \multicolumn{1}{c}{nDCG@5} & \multicolumn{1}{c}{$\Delta(\%)$} & \multicolumn{1}{c}{MRR} & \multicolumn{1}{c}{$\Delta(\%)$} \\
\midrule
LinearCatRev & 0.4897  & \multicolumn{1}{c}{-} & 0.3213 & \multicolumn{1}{c}{-} & 0.6284 & \multicolumn{1}{c}{-} \\
GeoSoCa & 0.4207* & $-$14.09 & 0.2958 & $-$7.94 & 0.6497 & 3.39\\
nDTF & 0.4172* & $-$14.80 & 0.2663* & $-$17.12 & 0.6167 & $-$1.86\\
UT-ML & 0.5138 & 4.92 & 0.3357 & 4.48 & 0.6389 & 1.67\\
UT-CRF & 0.5138 & 4.92 & 0.3410 & 6.13 & 0.6765 & 7.65\\
UT-SVM & 0.5207 & 6.33 & 0.3389 & 5.48 & 0.6510 & 3.60\\
PK-Boosting & \textbf{0.5310} & \textbf{8.43} & \textbf{0.3526*} & \textbf{9.74} & \textbf{0.6800} & \textbf{8.21} \\
\bottomrule
\end{tabular}
\end{table}

\partitle{Impact of Different Learning to Rank Techniques.} 
In this experiment we aim to show how the recommendation effectiveness is affected by applying different learning to rank techniques to combine the scores. Tables~\ref{tb:ltr2015} and \ref{tb:ltr2016} report P@5 applying different learning to rank techniques for TREC-CS 2015 and TREC-CS 2016 respectively. We report the performance for UT-ML, UT-CRF, UT-SVM, and PK-Boosting. As we can see, ListNet in many cases outperforms other learning to rank techniques. More specifically, for TREC-CS 2015, ListNet exhibits the best performance for all models except for UT-ML. It is very interesting that RankNet exhibits the best performance for UT-ML, and both ListNet and RankNet are based on artificial neural networks. As for TREC-CS 2016, RankNet performs better for UT-ML and UT-SVM while ListNet performs better for other models. As we can observe, applying different learning to rank techniques can potentially have a big impact on recommendation results. Therefore, it is critical to apply the best technique for the scores.

\begin{table}
\centering
\caption{Effect on P@5 for different learning to rank techniques in TREC-CS 2015. Bold values denote the best learning to rank technique per model.}
\label{tb:ltr2015}
\begin{tabular}{l@{\quad}c@{\quad}c@{\quad}c@{\quad}c}
\toprule
 & UT-ML & UT-CRF & UT-SVM & PK-Boosting \\
\midrule
MART & 0.5911 & 0.6008 & 0.5958 & 0.6010 \\
RankNet & \textbf{0.6224} & 0.6190 & 0.6155 & 0.6190 \\ 
RankBoost & 0.6030 & 0.6086 & 0.6088 & 0.6146 \\
AdaRank & 0.6028 & 0.6121 & 0.6117 & 0.5893 \\
CoordinateAscent & .06115 & 0.5858 & 0.5918 & 0.5997 \\
LambdaMART & 0.6022 & 0.6077 & 0.6061 & 0.6135 \\
ListNet & 0.6069 & \textbf{0.6249} & \textbf{0.6219} & \textbf{0.6259} \\
RandomForests & 0.5836 & 0.5966 & 0.5920 & 0.5963 \\
\bottomrule
\end{tabular}

\bigskip

\centering
\caption{Effect on P@5 for different learning to rank techniques in TREC-CS 2016.}
\label{tb:ltr2016}
\begin{tabular}{l@{\quad}c@{\quad}c@{\quad}c@{\quad}c}
\toprule
 & UT-ML & UT-CRF & UT-SVM & PK-Boosting \\
\midrule
MART & 0.4653 & 0.4103 & 0.3931 & 0.4483 \\
RankNet & \textbf{0.5138} & 0.5103 & \textbf{0.5237} & 0.5103 \\ 
RankBoost & 0.3414 & 0.4241 & 0.4345 & 0.4586 \\
AdaRank & 0.3414 & 0.3414 & 0.3414 & 0.3414 \\
CoordinateAscent & 0.5021 & 0.4931 & 0.4931 & 0.5000 \\
LambdaMART & 0.3793 & 0.3931 & 0.3793 & 0.4931 \\
ListNet & 0.5103 & \textbf{0.5138} & 0.5103 & \textbf{0.5310} \\
RandomForests & 0.4207 & 0.4069 & 0.4345 & 0.4310 \\
\bottomrule
\end{tabular}
\end{table}

\partitle{Impact of Using Information from Multiple LBSNs.} 
Tables \ref{tb:sources2015} and \ref{tb:sources2016} evaluate the performance of the examined models before and after removing information from each LBSN. In this set of experiments, we also report the relative performance drop of different models when using information from the two different LBSNs. As we can see in almost all cases, when a source of information is removed from the model, we observe a drop in the performance. The average drop for TREC-CS 2015 is $-4.90\%$ and for TREC-CS 2016 is $-6.00\%$ which confirms the effectiveness of exploiting information from different LBSNs. This indicates that using multimodal information from different LBSNs is a key to improve POI recommendation. For all different runs, the best performing method is the proposed PK-Boosting, that uses a combination of information from both LBSNs. 

\begin{table}
\centering
\caption{Performance evaluation after removing information provided by Foursquare (F) and Yelp (Y) in the TREC-CS 2015 dataset. The superscript * denotes significant differences compared to the performance each model has when using information from the two different LBSNs. $\Delta$ values ($\%$) express the relative drop, compared to the performance each model has when using information from the two different LBSNs. (Average drop$=-4.90\%$)}
\label{tb:sources2015}
\begin{tabular}{l@{\quad}l@{\quad}l@{\quad}l@{\quad}r@{\quad}l@{\quad}r@{\quad}l@{\quad}r}
\toprule
 & F & Y & \multicolumn{1}{c}{P@5} & \multicolumn{1}{c}{$\Delta(\%)$} & \multicolumn{1}{c}{nDCG@5} & \multicolumn{1}{c}{$\Delta(\%)$} & \multicolumn{1}{c}{MRR} & \multicolumn{1}{c}{$\Delta(\%)$} \\
 \midrule
LinearCatRev & \cmark & \cmark & 0.5858  & \multicolumn{1}{c}{-} & 0.6055 & \multicolumn{1}{c}{-} & 0.7404 & \multicolumn{1}{c}{-} \\
 & \cmark & \xmark & 0.5649  & $-$3.57 & 0.5860 & $-$3.22 & 0.7263 & $-$1.90\\
 & \xmark & \cmark & 0.5697  & $-$2.75 & 0.5917 & $-$2.28 & 0.7341 & $-$0.85\\
\midrule
UT-ML & \cmark & \cmark & 0.6224 & \multicolumn{1}{c}{-} & 0.6320 & \multicolumn{1}{c}{-} & 0.7496 & \multicolumn{1}{c}{-} \\
 & \cmark & \xmark & 0.5288* & $-$15.04 & 0.5307* & $-$16.03 & 0.6487* & $-$13.46 \\
 & \xmark & \cmark & 0.5787* & $-$7.02 & 0.5746* & $-$9.08 & 0.6833* & $-$8.84 \\
\midrule
UT-CRF & \cmark & \cmark & 0.6249& \multicolumn{1}{c}{-} & 0.6285 & \multicolumn{1}{c}{-} & 0.7434 & \multicolumn{1}{c}{-} \\
 & \cmark & \xmark & 0.5960* & $-$8.95 & 0.5930* & $-$5.65 & 0.7301& $-$1.79 \\
 & \xmark & \cmark & 0.6055& $-$3.10 & 0.6238&  $-$0.75 & 0.7503& 0.93 \\
\midrule
UT-SVM & \cmark & \cmark & 0.6219 & \multicolumn{1}{c}{-} & 0.6339 & \multicolumn{1}{c}{-} & 0.7553 & \multicolumn{1}{c}{-} \\
 & \cmark & \xmark & 0.5728* & $-$7.90 & 0.5921* & $-$6.59 & 0.7388& $-$2.18\\
 & \xmark & \cmark & 0.6129& $-$1.45 & 0.6250& $-$1.40 & 0.7497& $-$0.74\\
\midrule
PK-Boosting & \cmark & \cmark & \textbf{0.6259} & \multicolumn{1}{c}{-} & \textbf{0.6409} & \multicolumn{1}{c}{-} & \textbf{0.7704} & \multicolumn{1}{c}{-} \\
 & \cmark & \xmark & 0.5731* & $-$8.44 & 0.6010* & $-$6.23 & 0.7602 & $-$1.32 \\
 & \xmark & \cmark & 0.6044& $-$3.44 & 0.6227& $-$2.84 & 0.7613& $-$1.18 \\

\bottomrule
\end{tabular}

\bigskip

\centering
\caption{Performance evaluation after removing information provided by Foursquare (F) and Yelp (Y) in the TREC-CS 2016 dataset. The superscript * denotes significant differences compared to the performance each model has when using information from the two different LBSNs. (Average drop$=-6.00\%$)} 
\label{tb:sources2016}
\begin{tabular}{l@{\quad}l@{\quad}l@{\quad}l@{\quad}r@{\quad}l@{\quad}r@{\quad}l@{\quad}r}
\toprule
 & F & Y & \multicolumn{1}{c}{P@5} & \multicolumn{1}{c}{$\Delta(\%)$} & \multicolumn{1}{c}{nDCG@5} & \multicolumn{1}{c}{$\Delta(\%)$} & \multicolumn{1}{c}{MRR} & \multicolumn{1}{c}{$\Delta(\%)$} \\
 \midrule
LinearCatRev & \cmark & \cmark & 0.4897  & \multicolumn{1}{c}{-} & 0.3213 & \multicolumn{1}{c}{-} & 0.6284 & \multicolumn{1}{c}{-} \\
 & \cmark & \xmark & 0.4172* & $-$14.49 & 0.2705* & $-$15.81 &0.6222& $-$0.99\\
 & \xmark & \cmark & 0.4759& $-$2.46& 0.3072& $-$4.39& 0.6032& $-$4.01 \\
\midrule
UT-ML & \cmark & \cmark & 0.5138 & \multicolumn{1}{c}{-} & 0.3357 & \multicolumn{1}{c}{-} & 0.6389 & \multicolumn{1}{c}{-}\\
 & \cmark & \xmark & 0.4862 & $-$5.37 & 0.3079 & $-$8.28 & 0.6038 & $-$5.49 \\
 & \xmark & \cmark & 0.5034& $-$2.02 & 0.3313& $-$1.31 & 0.6393& 0.06\\
\midrule
UT-CRF & \cmark & \cmark & 0.5138 & \multicolumn{1}{c}{-} & 0.3410 & \multicolumn{1}{c}{-} & 0.6765 & \multicolumn{1}{c}{-} \\
 & \cmark & \xmark & 0.5069& $-$1.34 & 0.3336& $-$2.17 & 0.6531& $-$3.46\\
 & \xmark & \cmark & 0.4793& $-$6.71 & 0.3133& $-$8.12 & 0.6268& $-$7.35\\
\midrule
UT-SVM & \cmark & \cmark & 0.5207 & \multicolumn{1}{c}{-} & 0.3389 & \multicolumn{1}{c}{-} & 0.6510 & \multicolumn{1}{c}{-} \\
 & \cmark & \xmark & 0.4724 & $-$9.28 & 0.3057* & $-$9.80 & 0.6260& $-$3.84\\
 & \xmark & \cmark & 0.4793& $-$7.95 & 0.3158& $-$6.82 & 0.6512& 0.03\\
\midrule
PK-Boosting & \cmark & \cmark & \textbf{0.5310} & \multicolumn{1}{c}{-} & \textbf{0.3526} & \multicolumn{1}{c}{-} & \textbf{0.6800} & \multicolumn{1}{c}{-} \\
 & \cmark & \xmark & 0.4793* & $-$9.74 & 0.3210& $-$8.39 & 0.6542& $-$3.79\\
 & \xmark & \cmark & 0.4759* & $-$10.38 & 0.3177* & $-$9.90 & 0.6354&$-$6.56 \\

\bottomrule
\end{tabular}
\end{table}

\partitle{Impact of Using Different Scores.}
In this experiment, we try to demonstrate the effectiveness of each score. We remove each score and analyze our model's performance without it (but we do not remove more than one score at a time). The results are reported in Table \ref{tb:comp_removal}. 
The first line (\textit{All}) shows the results for P@5, nDCG@5, and MRR using all scores. The second line ($-S_{cat}$) shows the results without the location categories, and so on for the other lines. 

The results show a decrease of the model's performance after removing each of the scores exhibiting an average relative drop of $-4.31$\%. It indicates that our system is able to capture different aspects of information and combine them to create a better personalized ranking model for POI recommendation.
The $S_{cat}$ score models the types of locations a user is interested in visiting, while $S_{rev}$ models the reasons the user likes/dislikes different locations belonging to the same category. $S_{key}$ tries to incorporate the 
most important keywords extracted from the reviews and to describe a location and its characteristics. $S_{boost}$ boosts the most important keywords that interest a user and the contextual relevance is measured by $S_{cxt}$.
Our model exhibits its largest decrease in performance when $S_{rev}$ is removed from the model. This suggests that the review-based score is the most important score in our model. We think this is because it captures users' opinions. In fact, it is crucial to realize why a user rates two locations in the same category differently.

\begin{table}
    \centering
    \caption{Performance of PK-Boosting using all the scores (\textit{All}) and after removing each score at a time. 
    The superscript * denotes significant differences compared to the model using all scores (\textit{All}).
    Percentages in bold represent the highest decrease in performance when the corresponding score is removed (Average relative drop = $-4.31$\%).}
    \label{tb:comp_removal}
    \begin{tabular}{l @{\quad}l @{\quad}r @{\quad}l @{\quad}r @{\quad}l @{\quad}r}
        \toprule
                     & \multicolumn{1}{c}{P@5} & \multicolumn{1}{c}{$\Delta(\%)$} & \multicolumn{1}{c}{nDCG@5} & \multicolumn{1}{c}{$\Delta(\%)$} & \multicolumn{1}{c}{MRR} & \multicolumn{1}{c}{$\Delta(\%)$} \\
        \midrule
           ~~~~~~\textit{All}                        & 0.6259 & \multicolumn{1}{c}{-} & 0.6409 & \multicolumn{1}{c}{-} & 0.7704 & \multicolumn{1}{c}{-}\\ 
         \textit{$- S_{cat}$}      & 0.6009* & $-$3.99 & 0.6124* & $-$4.45 & \textbf{0.7324}* & \textbf{$-$4.93}\\ 
         \textit{$- S_{rev}$}      & \textbf{0.5555}* & \textbf{$-$11.25} & \textbf{0.5837}* & \textbf{$-$8.92} & 0.7383* & $-$4.17 \\ 
         \textit{$-S_{key}$}     & 0.6009* & $-$3.99 & 0.6113 & $-$3.35 & 0.7443 & $-$3.10 \\ 
         \textit{$-S_{boost}$}      & 0.6190 & $-$1.10 & 0.6312 & $-$1.51 & 0.7610 & $-$1.22 \\
         \textit{$-S_{cxt}$}      & 0.5962* & $-$4.75 & 0.6126* & $-$4.42 & 0.7437 & $-$3.47 \\         
         \bottomrule
    \end{tabular}
\end{table}

\begin{figure} \centering
\begin{tabular}{cc}
\includegraphics[width=0.48\textwidth]{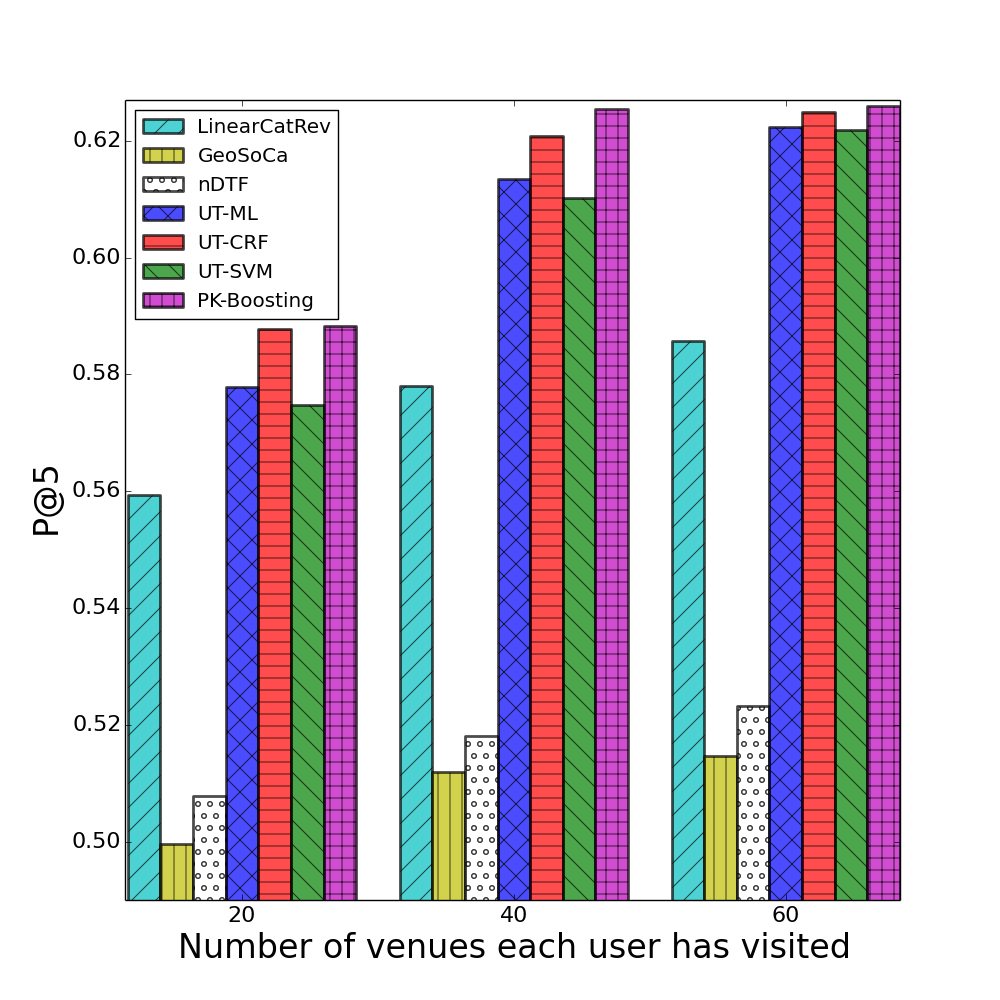} & \includegraphics[width=0.48\textwidth]{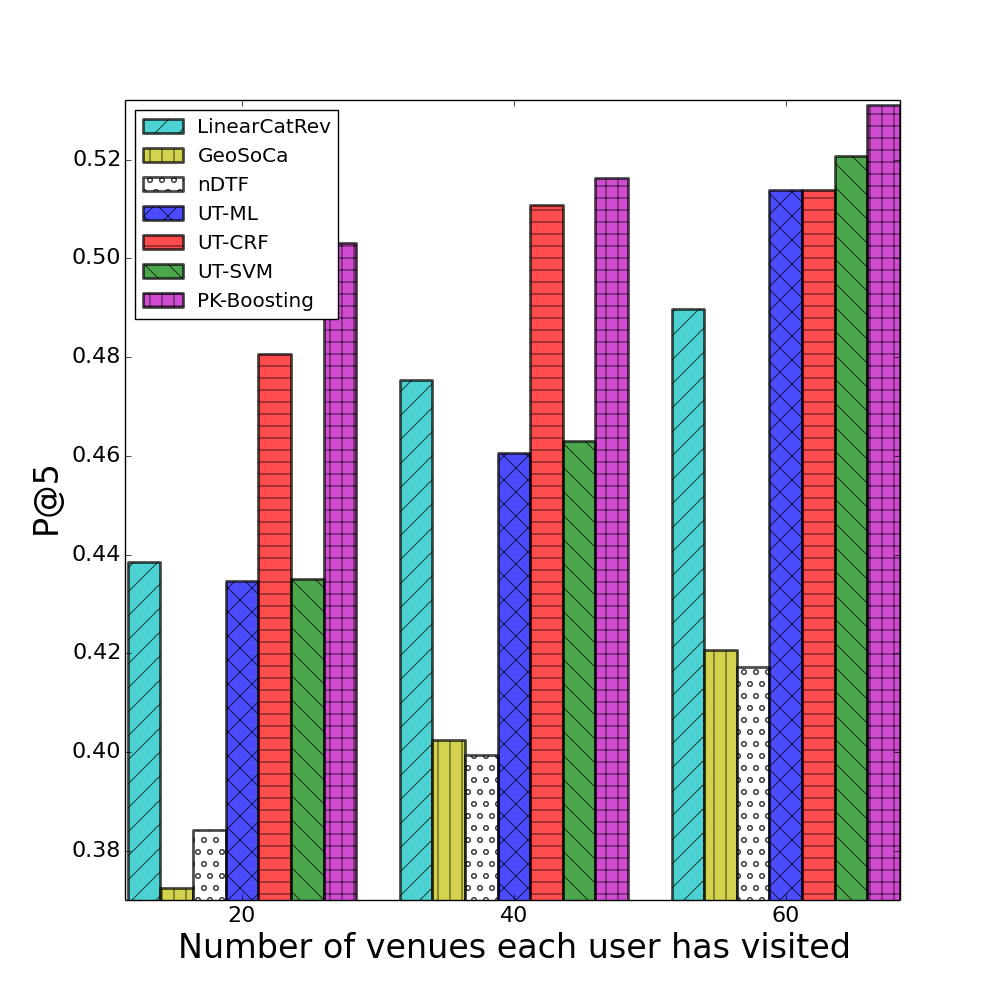}\\
(a) TREC-CS 2015  & (b) TREC-CS 2016 \\
\end{tabular}
\caption{Effect on P@5 by varying the number of locations that each user has visited for (a) TREC-CS 2015 and (b) TREC-CS 2016.} \label{fg:no_of_venues}
\end{figure}

\partitle{Impact of Number of Visited Locations.} 
We report P@5 of all models on TREC-CS 2015 and TREC-CS 2016 in Figure \ref{fg:no_of_venues}. In this set of experiments, we vary the number of locations to find the mapping between the taste keywords and the user tags. We calculate the scores of Section~\ref{sc:suggestion} with different number of locations and train the ranking model. Figure \ref{fg:no_of_venues} shows that PK-Boosting is the winning method when compared with other models for all different number of locations. 
This result indicates that PK-Boosting is more robust when the training set is smaller, whereas the prediction models ML and SVM are not very well trained using such a small data and their performance gets worse.

\partitle{Impact of Visiting POIs from a Single City vs. Two Cities.}
In this experiment, we intend to see how the number of visited locations from one single city affects the performance of our model as opposed to the same number of locations from different cities. In order to do that, we consider at maximum 2 cities, and we train our model using 10, 15, ..., 60 of them for each user as their history of preferences. To make sure that the order of selected locations does not affect our experiment, we shuffle the list of previously visited locations in two ways: 1) we make sure that the first 30 locations are from a city and the second 30 are from the other city; 2) we shuffle the order of visited locations for each city and interleave them. For example, $v_1$ would be a location from City1, $v_2$ a location from City2, $v_3$ a location from City1, and so on. We conduct this experiment with 5 differently shuffled lists and report the average of the results.

The first ordering method ensures that the first half of the locations, visited by a user, are from a particular city. The second ordering, on the other hand, makes sure that for a given number of visited locations $n$, $n/2$ of them are locations from City1 and $n/2$ are from City2 . We intend to examine how our model performs when we have information about users from only one city as opposed to two cities. 
Moreover, in cases where we have a low number of visited locations in the user's history, it is interesting to see how our model performs in both scenarios (i.e., all locations from a single city vs. from multiple cities).

\begin{figure}
    \centering
    \includegraphics[width=0.50\textwidth]{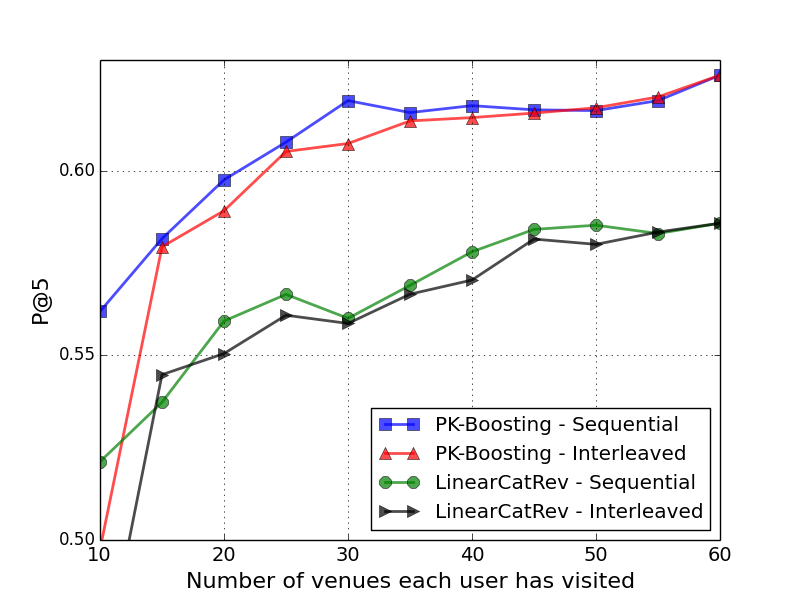}
    \caption{Our model's performance in terms of P@5 with different number of locations as users' history of preferences compared to LinearCatRev. We have chose the order of locations in two different manners. \textit{Sequential}: the first 30 locations are from one single city, the second 30 are from another city, \textit{Interleaved}: the list of locations is interleaved based on their cities.}
    \label{fg:tr}
\end{figure}

Figure \ref{fg:tr} demonstrates our system performance in terms of P@5 with different number of locations in the user's history compared to LinearCatRev. As we can see in the figure, the model shows a large improvement up to the first 30 locations, decreasing in size after we add 40 locations in both orderings. However, it is interesting to see that the sequential order always performs better than the interleaved one. This difference is more evident when the number of locations is smaller than 20. It suggests that when we have limited number of locations as the user's history, it is better to have them all about the same city. This can be observed when there are 30 locations and all from one single city (denoted as \textit{Sequential}), we get a much better performance as compared to training the model using 30 locations with half from one city and the other half from another one (denoted as \textit{Interleaved}).

\subsection{Dimensionality Reduction}
In this experiment we compare our personalized keyword boosting method with the well-known dimensionality reduction method PCA. Since  keyword boosting is a kind of personalized dimensionality reduction, we choose to compare our method with PCA. As Tables~\ref{tb:dim_2015} and \ref{tb:dim_2016} show, our personalized keyword boosting method is able to beat PCA with respect to recommendation effectiveness in terms of both $P@k$ and $nDCG@k$. It suggests that the proposed probabilistic model is able to effectively reduce the dimensionality of location keywords taking into account user personal preferences as well as interests. In fact, in TREC-CS 2015 the average location keyword per user is 277 and our proposed approach is able to reduce it to 41 ($-$\%85), whereas PCA reduces it to 25 ($-$\%91). Moreover, the average location keywords per user in TREC-CS 2016 is 302 and PK-Boosting reduces it to 105 ($-$\%65) compared with PCA reducing it to 16 ($-$\%95). 

It is worth noting that PCA produces $\text{min}(n, m-1)$ principal components, where $n$ is the number of data samples and $m$ is the number of data dimensions. Since in both datasets $n \ll m$, the number of principal components is bounded by $n$ and thus PCA reduces data dimensionality more than PK-Boosting. This difference for TREC-CS 2016 is even bigger due to the fact that users have smaller number of previously visited locations. 
It is interesting to note the difference between the dimensionality reduction that PK-Boosting exhibits on TREC-CS 2015 and TREC-CS 2016 (i.e., 41 vs. 105). 
This is due the difference between the average number of user tags in the two datasets. 
As we can see in Table~\ref{tb:ds-stats}, the average user tag per user for TREC-CS 2015 is 1.46 as compared with 3.61 for TREC-CS 2016. Since PK-Boosting personalizes dimensionality reduction problem according to user tags, the average number of user tags can potentially have an impact on the average number of reduced dimensions. The results of Tables~\ref{tb:dim_2015} and \ref{tb:dim_2016} show that PK-Boosting outperforms PCA in terms of recommendation effectiveness even though PCA is able to reduce more dimensions. 
This suggests that incorporating personal information for dimensionality reduction is effective and hence PK-Boosting performs better.

\begin{table}
\centering
\caption{Performance comparison on TREC-CS 2015 on dimensionality reduction. The superscript * denotes significant differences compared to LinearCatRev.}
\label{tb:dim_2015}
\begin{tabular}{l@{\quad}l@{\quad}l@{\quad}r@{\quad}l@{\quad}r@{\quad}l@{\quad}r}
\toprule
 & Avg. Dim. & \multicolumn{1}{c}{P@5} & \multicolumn{1}{c}{$\Delta(\%)$} & \multicolumn{1}{c}{nDCG@5} & \multicolumn{1}{c}{$\Delta(\%)$} & \multicolumn{1}{c}{MRR} & \multicolumn{1}{c}{$\Delta(\%)$} \\
\midrule
LinearCatRev & 277 & 0.5858  & \multicolumn{1}{c}{-} & 0.6055 & \multicolumn{1}{c}{-} & 0.7404 & \multicolumn{1}{c}{-} \\
PK-PCA & 25 & 0.6030* & 4.37 & 0.6157 & 3.07 & 0.7366 & $-$0.32 \\
PK-Boosting & 41 & \textbf{0.6259}* & \textbf{6.85} & \textbf{0.6409}* & \textbf{5.85} & \textbf{0.7704}* & \textbf{4.05} \\
\bottomrule
\end{tabular}

\bigskip

\centering
\caption{Performance comparison on TREC-CS 2016 on dimensionality reduction. The superscript * denotes significant differences compared to LinearCatRev.}
\label{tb:dim_2016}
\begin{tabular}{l@{\quad}l@{\quad}l@{\quad}r@{\quad}l@{\quad}r@{\quad}l@{\quad}r}
\toprule
 & Avg. Dim. & \multicolumn{1}{c}{P@5} & \multicolumn{1}{c}{$\Delta(\%)$} & \multicolumn{1}{c}{nDCG@5} & \multicolumn{1}{c}{$\Delta(\%)$} & \multicolumn{1}{c}{MRR} & \multicolumn{1}{c}{$\Delta(\%)$} \\
\midrule
LinearCatRev & 302 & 0.4897  & \multicolumn{1}{c}{-} & 0.3213 & \multicolumn{1}{c}{-} & 0.6284 & \multicolumn{1}{c}{-} \\
PK-PCA & 16 & 0.5106 & 4.21 & 0.3406 & 6.02 & 0.6424 & 2.23\\
PK-Boosting & 105 & \textbf{0.5310} & \textbf{8.43} & \textbf{0.3526}* & \textbf{9.74} & \textbf{0.6800} & \textbf{8.21} \\
\bottomrule
\end{tabular}
\end{table}

\subsection{User Tag Prediction}
In this experiment we evaluate the effectiveness of different user tag prediction methods. The aim of this experiment is to show how effective user tag prediction is in terms of user tag prediction accuracy. In previous experiments we showed how user tag prediction can improve the overall recommendation effectiveness; however, it is crucial to know how effective is the prediction model so that we can further analyze and improve the prediction accuracy in order to achieve better recommendation. Table~\ref{tab-pred} reports the performance of different user tag prediction models. Note that CRF and SVM-based taggers are trained using the same feature set for fair comparison. As we can see in this table the SVM based model is able to beat all other models. In fact, the SVM based model benefits highly from the features that are extracted using the proposed mapping and therefore it can beat ML. 

\begin{table}
    \centering
    \caption{Performance comparison of user tag prediction models.}
    \label{tab-pred}
    \begin{tabular}{l @{\quad}c @{\quad}c @{\quad}c}
\toprule
                     & Precision & Recall & F-Measure \\
        \midrule
         ML		& 0.3982 & 0.2421 & 0.3011 \\
         SVM-Based     & \textbf{0.7923} & \textbf{0.8110} & \textbf{0.8016} \\ 
         CRF   & 0.7646 & 0.7573 & 0.7609  \\ 
         \bottomrule
  
    \end{tabular}
\end{table}

\section{Related Work}\label{RelatedWork}
In this section, we review some of the existing related works on POI recommendation, and context-aware POI recommendation.

\partitle{POI Recommendation.}
Recommender systems play an important role in satisfying users' expectations for many online services such as e-commerce, LBSN and social network websites.
CF-based approaches are based on the core idea that users with similar behavioral history tend to act similarly in the future~\cite{DBLP:journals/cacm/GoldbergNOT92}. A large body of research has been done following this idea~\cite{DBLP:conf/recsys/GriesnerAN15,DBLP:journals/tois/YinCSHC14,DBLP:journals/tist/ZhangDCLZ13,DBLP:conf/cikm/FerenceYL13}.
CF-based approaches often suffer from data sparsity since there are a lot of available locations, and a single user can visit only a few of them. As a consequence, the user-item matrix of CF becomes very sparse, leading to poor performance of recommender systems in cases that there is no significant association between users and items. 
Many studies have tried to address the data sparsity problem of CF by incorporating additional information into the model~\cite{DBLP:conf/sigir/YeYLL11,DBLP:conf/sigir/YuanCMSM13}. More specifically, \citet{DBLP:conf/sigir/YeYLL11} argued that users' check-in behavior is affected by the spatial influence of locations and proposed a unified location recommender system incorporating spatial and social influence to address the data sparsity problem. \citet{DBLP:conf/sigir/YuanCMSM13}, on the other hand, proposed a time-aware collaborative filtering approach. More specifically, they recommended locations to users at a particular time of the day by mining historical check-ins of users in LBSNs. \citet{DBLP:journals/tois/YinCSHC14} proposed a model which captures user interests as well as local preferences to recommend locations or events to users when they are visiting a new city. \citet{DBLP:conf/cikm/FerenceYL13} took into consideration user preference, geographical proximity, and social influences for POI recommendation. \citet{DBLP:conf/recsys/GriesnerAN15} also proposed an approach integrating temporal and geographic influences into matrix factorization. 

Other works follow a review-based strategy, constructing rich user profiles based on their reviews~\cite{yang2015opinions,DBLP:journals/tist/ZhangDCLZ13}. Reviews reveal the underlying reasons of users' ratings related to a particular location. In fact, as argued by \citet{DBLP:journals/umuai/ChenCW15}, online reviews significantly aid a system to deal with the data sparsity problem. \citet{DBLP:journals/tist/ZhangDCLZ13} fused virtual ratings derived from online reviews into CF. 
\citet{yang2015opinions} created rich user profiles aggregating online reviews from other users and measured the similarity between a new location and a user profile. We exploit online reviews similarly, however, we employ more sophisticated machine learning models to learn users' preferences and opinions. \citet{DBLP:journals/corr/abs-1803-08354} studied various strategies for reducing the number of reviews in users' profiles while maintaining the model's efficiency. 

\partitle{Context-Aware POI Recommendation.}
Another line of research tries to leverage context to enhance the performance of a recommender system. Context-aware recommendation has been categorized into three types~\cite{DBLP:reference/rsh/AdomaviciusT11}: (1) \textit{pre-filtering}: data selection is done based on context; (2) \textit{post-filtering}: recommendation is done using a traditional approach and context is used to filter them; (3) \textit{contextual modeling}: contextual information is incorporated into the model. Our work aims at modeling the contextual information by re-ranking the recommendations.
\citet{DBLP:journals/tois/AdomaviciusSST05} proposed a multidimensional context pre-filtering model based on the online analytical processing for decision support.
\citet{DBLP:conf/uic/ParkHC07} computed a weighted sum of the conditional probabilities of restaurants' attribute values. They automatically detected users' physical contexts such as the time of the day, the position, and the weather and used a Bayesian network for expressing their probabilistic influences. \citet{DBLP:conf/recsys/LeviMDT12a} developed a weighted context-aware recommendation algorithm to address the cold start problem for hotel recommendation. More specifically, they defined context groups based on hotel reviews and followed a user's preferences in trip intent and hotel aspects as well as the user's similarity with other users (e.g., nationality). Other works focused on time as context~\cite{DBLP:conf/sigir/YuanCMSM13, DBLP:conf/recsys/GaoTHL13, DBLP:conf/cikm/DeveaudAMO15,DBLP:journals/tist/FangXHM16}. \citet{DBLP:conf/recsys/GaoTHL13} developed a time-aware recommendation model. \citet{DBLP:journals/tist/FangXHM16} proposed a model which takes into account both spatial and temporal context to address the data sparsity problem. \citet{DBLP:conf/cikm/DeveaudAMO15} modeled locations popularity in the immediate future utilizing time series. They leveraged the model to make time-aware POI recommendation. \citet{DBLP:conf/sac/AliannejadiMC17} recommended POIs while considering, as users' contexts, the traveling group as well as the season in which the trip occurred.
\citet{DBLP:journals/ia/BraunhoferER14} used various complex contextual factors such as budget, companion, and crowdedness to overcome the cold start problem. They developed an active learning strategy and a context-aware recommendation algorithm using an extended matrix factorization model.

\partitle{TREC Contextual Suggestion.}
The TREC-CS track~\cite{hashemi2016overview} aimed to encourage research on context-aware POI recommendation. In fact, the task was to produce a ranked list of locations for each user in a new city, given the user's context and history of preferences in 1-2 other cities. The contextual dimensions were the trip duration, the season, the trip type, and the type of group with whom the user was traveling. These contextual dimensions were introduced in TREC-CS 2015. Since then, among the top runs, few approaches tried to leverage such information.
\citet{Arampatzis:2017:SPV:3146384.3125620} studied the performance of various content-based, collaborative, and hybrid fusion methods on TREC-CS and they found that content-based methods performed best among these methods.
\citet{DBLP:conf/trec/Yang015} introduced some handcrafted rules for filtering locations based on their appropriateness to a user's current context. According to them, applying such filters degrades the performance of the system. Hence, we conclude that contextual appropriateness is not a simple problem of applying some deterministic rules to filter locations. 
\citet{DBLP:conf/clef/ManotumruksaMO16} introduced a set of temporal, term-based, and categorical features to train a set of classifiers to predict contextually appropriate locations. In our opinion, such features are not general enough to be applied to similar problems. Moreover, similar to our work, they collected the contextual relevance dataset using crowdsourcing, however, since they asked the workers to assess the appropriateness of a particular location to a given user's context, this could result in biased assessments. In contrast, we attempt to build a contextual relevance dataset that is not biased and can be used for similar problems. 
\section{Conclusions}\label{Conclusions}
In this paper, we presented a probabilistic model to find the mapping between user tags and location taste keywords. This mapping enabled us to exploit various directions to address the data sparsity problem for POI recommendation. In particular, we followed two directions: 1) a PK-Boosting model to reduce the dimensionality of location taste keywords and 2) three models to predict user tags for a new location, as alternatives to PK-Boosting.
Moreover, we described how to incorporate the new information into POI recommendation, calculating different scores from information from multiple LBSNs. In addition, we also created a dataset to measure the contextual appropriateness of locations and explained how we used the dataset to improve our model. 
Following learning to rank techniques, the final POI recommendation ranking is obtained based on the computed scores.
The experimental results on two TREC collections demonstrate that our method outperforms state-of-the-art strategies. This confirms that the proposed approach, PK-Boosting, addresses the data sparsity problem capturing user preferences accurately. 

As future work, we plan to extend our contextual model to capture the time dimension and perform time-aware POI recommendation~\cite{DBLP:conf/kdd/LiuLLQX16}.

\begin{acks}
 This work was partially supported by the \grantsponsor{}{Swiss National Science Foundation (SNSF)}{http://www.snf.ch/en/Pages/default.aspx} under the project ``Relevance Criteria Combination for Mobile IR (RelMobIR).''
\end{acks}

\bibliographystyle{ACM-Reference-Format}
\bibliography{main}

\end{document}